\documentclass[aps,preprint,showpacs,superscriptaddress,groupedaddress,numbers,sort&compress]{revtex4}  % for double-spaced preprint

\usepackage{graphicx}  % needed for figures
\usepackage{dcolumn}   % needed for some tables
\usepackage{bm}        % for math
\usepackage{amssymb}   % for math
\usepackage{slashed}   % for Dirac Slash
\usepackage{amsmath}   % for mutiline eqn
\usepackage{simplewick} % for contraction
\usepackage{verbatim}  % for multi-line comment
\usepackage{color}
\usepackage{ifxetex}
\usepackage{epsf}
\usepackage[toc,page]{appendix}

\begin{document}

% The following information is for internal review, please remove them for submission
\widetext

% the following line is for submission, including submission to the arXiv!!
% \hspace{5.2in} \mbox{Fermilab-Pub-04/xxx-E}

\title{\large \bfseries \boldmath Factorization of Radiative Leptonic Decays of $B^-$ and $D^-$ Mesons Including the Soft Photon Region}
\author{Ji-Chong Yang} %\email{1120120049@mail.nankai.edu.cn}
\author{Mao-Zhi Yang} \email{yangmz@nankai.edu.cn}
\affiliation{School of Physics, Nankai University, Tianjin 300071, P.R. China}

\begin{abstract}
In this work, we study the radiative leptonic decays of $B^-$ and $D^-$ mesons using factorization approach. Factorization is proved to be valid explicitly at 1-loop level at any order of $O(\Lambda _{\rm QCD}\left/m_Q\right.)$. We consider the contribution in the soft photon region that $E_{\gamma} \sim \left. \Lambda^2 _{\rm QCD} /\right. m_Q$. The numerical results show that, the soft photon region is very important for both the $B$ and $D$ mesons. The branching ratios of $B\to \gamma e\nu _e$ is $5.21\times 10^{-6}$, which is about $3$ times of the result obtained by only considering the hard photon region $E_{\gamma}\sim m_Q$. And for the case of $D\to \gamma e\nu _e$, the result of the branching ratio is $1.92\times 10^{-5}$.
\end{abstract}

\maketitle

\section{\label{sec:level1}Introduction}

The study of the heavy meson decays is an important subject for understanding the high energy physics and standard model~\cite{review of B physics}. The rare decays of the pseudoscalar meson also provide a sensitive probe for new physics~\cite{importance of D rara decay}. In recent years, both experimental and theoretical studies have been improved greatly~\cite{experimatal data}. However, due to the quark confinement, one cannot probe the quarks directly in the experiments. The effect of hadronic bound-state have to be considered theoretically when treating hadronic decays. The hadronic bound-state effect is certainly highly none-perturbative in QCD. However the limitation in understanding and controlling the non-perturbative effects in QCD is so far still a problem.

Varies theoretical methods on how to deal with the non-perturbative effects have been developed. An important approach known as factorization~ \cite{DIS and Dran-Yan,Collinear Factorization} has been greatly developed to study the decay of the hadrons~\cite{Factorization in Decay}. In Ref.~\cite{SCET work neubert,Daniel Wyler work,SCET factorization}, the factorization is constructed using the soft-collinear effective theory (SCET)~\cite{SCET}. The idea of factorization is to absorb the infrared (IR) behavior into the wave-function of hadron, so the matrix element can be written as the convolution of wave-function and a hard kernel
\begin{equation}
\begin{split}
&F=\int dk \Phi(k)T_{\rm hard}(k)\\
\end{split}
\label{eq.0.1}
\end{equation}
The wave-function should be determined by non-perturbative methods.

The radiative leptonic decay of heavy mesons provides a good opportunity to study the factorization approach, where strong interaction is involved only in one hadronic external state. Many works has been done on this decay mode using factorization~\cite{TM Yan work,Genon work,Neubert work,Daniel Wyler work}. In these studies, the heavy quark is treated in the heavy quark effect theory (HQET)~\cite{heavy quark expansion}. In Ref.~\cite{Beneke work}, the factorization of radiative leptonic decay of $B$ meson is revisited with the next-to-leading logarithmic (NLL) contribution and $O(\Lambda _{\rm QCD}/m_Q)$ contribution at tree level considered. In Ref.~\cite{Our work} factorization is proved to order $O(\Lambda _{\rm QCD}\left. / \right. m_Q)$. In those works, the energy of photon in the hard region is treated where $E_{\gamma}\sim m_Q$. One cannot obtain the valid results in the region of soft photon because $O(\Lambda _{\rm QCD}\left. / E_{\gamma}\right.)^2$ contribution is neglected. The soft photon region is very important in the radiative decay both theoretically and numerically. Consider the tree level result for example, if the $\Lambda _{\rm QCD} \left. /E_{\gamma}\right.$ contribution in the decay amplitude is preserved, the branching ratios will be increased by
\begin{equation}
\begin{split}
&\int _{\Delta E_{\gamma}}^{\frac{m_Q}{2}}dE_{\gamma}\left(1+\frac{\Lambda _{\rm QCD}}{E_{\gamma}}\right)^2 \sim 1+\frac{\Lambda _{\rm QCD}}{\Delta E_{\gamma}}\\
\end{split}
\label{eq.0.2}
\end{equation}
where $\Delta E_{\gamma}$ is the cut on photon energy to regulate the soft photon. Using $\Lambda _{\rm QCD}=0.2\; {\rm GeV}$, and $\Delta E_{\gamma}$ can be as small as $\Delta E_{\gamma} \sim 0.01\;{\rm GeV}$, the result is possible to increase quite a lot no matter how heavy the meson is. However, this region is still absence in the previous works.

In this work, factorization of the radiative leptonic decays of heavy mesons is proved to be valid explicitly at 1-loop level at any order of $O(\Lambda _{\rm QCD}\left/m_Q\right.)$, where the soft photon region as $E_{\gamma} \sim \left. \Lambda^2 _{\rm QCD} /\right. m_Q$ is also considered. The numerical results show that, the soft photon region is very important for both $B$ and $D$ mesons. The branching ratios of $B\to \gamma e\nu _e$ can be increased to $5.21\times 10^{-6}$, which is about $3$ times of the result we obtained in Ref.~\cite{Our work}, where only hard photon region is seriously treated. For the case of $D\to \gamma e\nu _e$, the result is $1.92\times 10^{-5}$, which is close to Ref.~\cite{Our work}.

The remainder of the paper is organized as follows. In Sec.~\ref{sec:tree}, we present the factorization at tree level and discuss the kinematic of the radiative decay and the wave-function. In Sec.~\ref{sec:wavefunction}, the 1-loop corrections of the wave-function are discussed. The factorization at 1-loop order is presented in Sec.~\ref{sec:loop}. In Sec.~\ref{sec:resum}, we obtain the result in the soft photon region and hard photon region separately, and briefly discuss the resummation of the large logarithms. The numerical results are presented in Sec.~\ref{sec:numerical}. Finally Sec.~\ref{sec:summary} is a summary.

\section{\label{sec:tree}Tree Level result and kinematics}

The heavy pseudoscalar meson, $B$ or $D$ meson is constituted with a quark and an anti-quark, where one is heavy and the other is light. To study the factorization, we consider the state of a free quark and an anti-quark at first. The wave-function of the state can be defined as
\begin{equation}
\Phi (k_q,k_Q) = \int d^4xd^4y \exp (i k_q\cdot x)\exp (i k_Q\cdot y) \langle 0|\bar{q}(x)\left[x, y\right]Q(y)|\bar{q}Q\rangle
\label{eq.1.1}
\end{equation}
where $\left[x, y\right]$ denotes the Wilson line \cite{Wilson line}. And the matrix element is defined as
\begin{equation}
F= \langle\gamma|\bar{q}(x)P_L^{\mu}Q(y)|\bar{q}Q\rangle
\label{eq.1.2}
\end{equation}
The prove of factorization is to prove that, the matrix element can be written as the convolution of a wave-function $\Phi$ and a hard-scattering kernel $T$, where $T$ is infrared(IR) finite and independent of the external state.

\begin{figure}
\includegraphics[scale=0.9]{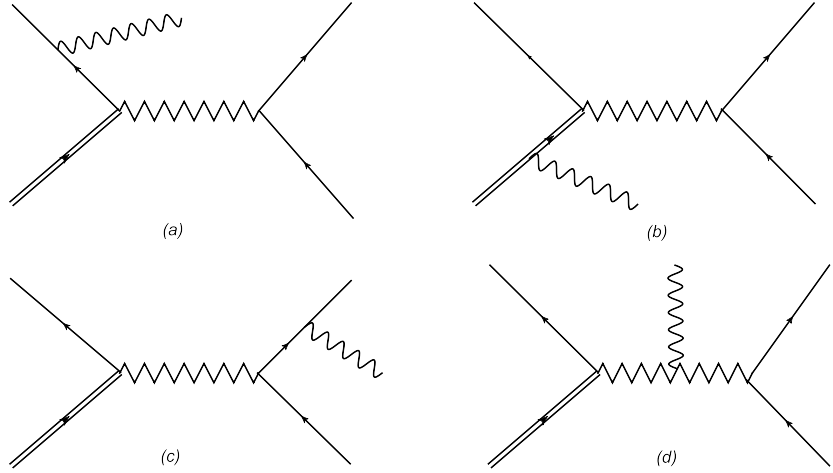}
\caption{\label{Fig1:feynman_tree} tree level amplitudes, the double line represents the heavy quark propagator but not the HQET propagator.}
\end{figure}

We start with the matrix elements at tree level. The Feynman diagrams of the radiative leptonic decay at tree level can be shown as Fig.~\ref{Fig1:feynman_tree}. The contribution of Fig.~\ref{Fig1:feynman_tree}.d is suppressed by a factor of $1\left/M_w^2\right.$, and can be neglected. The amplitudes of Figs.~\ref{Fig1:feynman_tree}.a, b and c can be written as
\begin{equation}
\begin{split}
&\mathcal{A}_a^{(0)}=\frac{i e_q G_F V_{Qq}}{\sqrt{2}}\bar{q}(p_{\bar{q}})\slashed \varepsilon _{\gamma }^*\frac{\slashed p_{\gamma }-\slashed p_q}{(p_{\gamma}-p_{\bar{q}})^2}P_L^{\mu } Q(p_Q)\left(lP_{L\mu}\bar{\nu}\right)\\
&\mathcal{A}_b^{(0)}=\frac{i e_Q G_F V_{Qq}}{\sqrt{2}}\bar{q}(p_{\bar{q}})P_L^{\mu }\frac{\slashed p_Q-\slashed p_{\gamma}+m_Q}{(p_Q-p_{\gamma})-m_Q^2}\slashed \varepsilon _{\gamma }^* Q(p_Q)\left(lP_{L\mu}\bar{\nu}\right)\\
&\mathcal{A}_c^{(0)}=\frac{- e G_F V_{Qq}}{\sqrt{2}}\bar{q}(p_{\bar{q}})P_L^{\mu } Q(p_Q)\left(l\slashed \varepsilon _{\gamma }^*\frac{i(\slashed p_{\gamma }+\slashed p_l+m_l)}{2\left(p_{\gamma }{\cdot}p_l\right)}P_{L\mu}\bar{\nu}\right)
\end{split}
\label{eq.1.3}
\end{equation}
where $p_{\bar{q}}$ and $p_Q$ are the momenta of the anti-quark $\bar{q}$ and quark $Q$, respectively. $p_{\gamma}$, $p_l$ and $p_{\nu}$ are the momenta of photon, lepton and neutrino, $\varepsilon _{\gamma}$ denotes the polarization vector of photon, and $P_L^{\mu}$ is defined as $\gamma^{\mu}(1-\gamma _5)$.

Lorentz invariant definition of the wave-function in coordinate space is
\begin{equation}
\Phi_{\alpha\beta}(x, y)=\langle 0|\bar{q}_{\alpha}(x)[x,y]Q_{\beta}(y)|\bar{q}^S(p_{\bar{q}}),Q^s(p_Q)\rangle
\label{eq.1.4}
\end{equation}
where $S$ and $s$ are spin labels of $\bar{q}$ and $Q$, respectively. Then we find
\begin{equation}
\begin{split}
&\Phi^{(0)}_{\alpha\beta}(k_{\bar{q}}, k_Q)=(2\pi)^4\delta^4(k_{\bar{q}}-p_{\bar{q}})(2\pi)^4\delta^4(k_Q-p_Q) \bar{v}_{\alpha}(p_{\bar{q}})u_{\beta}(p_Q)
\end{split}
\label{eq.1.5}
\end{equation}
And then the matrix element is
\begin{equation}
\begin{split}
&F^{(0)}=\int \frac{d^4k_{\bar{q}}}{(2\pi) ^4}\frac{d^4k_Q}{(2\pi)^4}\Phi^{(0)}(k_{\bar{q}}, k_Q)T^{(0)}(k_{\bar{q}}, k_Q)=\Phi^{(0)}\otimes T^{(0)}\\
\end{split}
\label{eq.1.6}
\end{equation}
with
\begin{equation}
\begin{split}
&F_a^{(0)}=e_q\slashed \varepsilon _{\gamma }^* \frac{\slashed p_{\gamma }-\slashed p_{\bar{q}}}{(p_{\gamma}-p_{\bar{q}})-m_q^2}P_L^{\mu }\\
&F_b^{(0)}=e_QP_L^{\mu }\frac{\slashed p_Q-\slashed p_{\gamma}+m_Q}{(p_Q-p_{\gamma})^2-m_Q^2}\slashed \varepsilon _{\gamma }^*\\
\end{split}
\label{eq.1.7}
\end{equation}

Using Eqs.~(\ref{eq.1.5}-\ref{eq.1.7}), we find there are a few different valid ways to define the tree level hard-scattering kernel, which fulfill the requirement of the factorization. For example, the one we used in Ref.~\cite{Our work} is
\begin{equation}
\begin{split}
&T_a^{(0)}=-e_q\frac{\slashed \varepsilon _{\gamma }^*\slashed p_{\gamma }-2 \varepsilon _{\gamma }^*\cdot k_q}{2p_{\gamma}\cdot k_{\bar{q}}}P_L^{\mu },\;\;\;\;\;T_b^{(0)}=-e_QP_L^{\mu }\frac{2 \varepsilon _{\gamma }^*\cdot k_Q-\slashed p_{\gamma}\slashed \varepsilon _{\gamma }^*}{2k_Q\cdot p_{\gamma}}
\end{split}
\label{eq.1.8}
\end{equation}
and another valid definition is:
\begin{equation}
\begin{split}
&T_a^{(0)}=e_q\slashed \varepsilon _{\gamma }^* \frac{\slashed p_{\gamma }-\slashed k_{\bar{q}}}{(p_{\gamma}-k_{\bar{q}})^2-m_q^2}P_L^{\mu },\;\;\;\;\;T_b^{(0)}=e_QP_L^{\mu }\frac{\slashed k_Q-\slashed p_{\gamma}+m_Q}{(k_Q-p_{\gamma})^2-m_Q^2}\slashed \varepsilon _{\gamma }^*
\end{split}
\label{eq.1.9}
\end{equation}
We find those two different $T^{(0)}$'s will lead to different results. Especially for the case that emitting a soft photon where $E_{\gamma}$ is small, using the definition in Eq.~(\ref{eq.1.8}), there will be IR divergence which can not be canceled, and the factorization will fail. However, if we use the definition of Eq.~(\ref{eq.1.9}), we find that, at 1-loop order, all IR divergence can be absorbed into the wave-function to all orders of $\Lambda _{\rm QCD}\left./\right.m_Q$. We will briefly discuss it in Sec.~III. To study the decay including the soft photon region, we choose the definition in Eq.~(\ref{eq.1.9}).

The amplitude $\mathcal{A}_c$ leads to another hard scattering kernel $T_c^{(0)}$. We find
\begin{equation}
\begin{split}
&T_c^{(0)}=-e\frac{\slashed p_{\gamma }\slashed \varepsilon _{\gamma }^*}{2p_{\gamma}\cdot p_l}P_L^{\mu }+eP_L^{\mu}\left(-\frac{\varepsilon \cdot p_l}{p_{\gamma }\cdot p_l}\right)
\end{split}
\label{eq.1.10}
\end{equation}

The kinematic feature of the soft photon region and hard photon region is very different. We work in the rest-frame of the meson, and we choose the frame such that the direction of the photon momentum is on the opposite z axis, so the momentum of the photon can be written as $p_{\gamma}=(E_{\gamma}, 0, 0, -E_{\gamma})$, with $0\le E_{\gamma} \le m_Q \left/ 2 \right.$. To consider the contribution when the photon is soft, we consider the region of the energy of photon $E_{\gamma}$ separately. In the soft photon region, we assume $E_{\gamma}\sim \Lambda _{\rm QCD}^2 \left. / m_Q\right.$, while in the hard photon region, we assume $E_{\gamma}\sim m_Q$.

When $E_{\gamma}\sim m_Q$, only part of $F_a^{(0)}$ is at the leading order of heavy quark expansion
\begin{equation}
\begin{split}
&F_{a\;\rm leading}^{(0)}=e_q\slashed \varepsilon _{\gamma }^* \frac{\slashed p_{\gamma }}{(p_{\gamma}-p_{\bar{q}})-m_q^2}P_L^{\mu}\sim O(1)
\end{split}
\label{eq.1.11}
\end{equation}
The others are at the order $O\left(\Lambda _{\rm QCD} /m_Q\right)$. Gauge invariance is kept order by order. When $E_{\gamma}\sim \Lambda _{\rm QCD}^2\left./m_Q\right.$, also part of $F_a^{(0)}$ is at the leading order, and
\begin{equation}
\begin{split}
&F_{a\;\rm leading}^{(0)}=e_q\slashed \varepsilon _{\gamma }^* \frac{-\slashed p_{\bar{q} }}{(p_{\gamma}-p_{\bar{q}})-m_q^2}P_L^{\mu}\sim O(\frac{m_Q}{\Lambda _{\rm QCD}})
\end{split}
\label{eq.1.12}
\end{equation}
\begin{comment}
To avoid confusion, we will always call the leading order as order $O(1)$. Notice that it is relatively $O(1)$ compared to other terms, in the soft photon case, the relatively $O(1)$ contribution is at order $O\left(m_Q/\Lambda _{\rm QCD}\right)$, and the relatively order $O\left(\Lambda _{\rm QCD}/m_Q\right)$ contribution is at order $O(1)$.
\end{comment}

With the help of $u_Q$, the $F_b^{(0)}$ can be written as
\begin{equation}
\begin{split}
&F_b^{(0)}=e_QP_L^{\mu }\frac{2p_Q\cdot \varepsilon -\slashed p_{\gamma}\slashed \varepsilon }{(p_Q-p_{\gamma})^2-m_Q^2}\\
\end{split}
\label{eq.1.13}
\end{equation}
so $F_b^{(0)}$ is found to be at sub-leading order because the polarization vector $\varepsilon$ dose not have the $0$-component.

Different from the hard photon case, in the soft photon region, gauge invariance is not kept at the leading order explicitly. In fact, part of $F_c^{(0)}$ becomes leading order. We find that gauge invariance is kept explicitly to the sub-leading order. So, for $E_{\gamma}$ is small, we should consider the sub-leading order contribution.

\section{\label{sec:wavefunction}The correction of wave-function at $O(\alpha _s)$ order}

The expansion of the decay amplitude can be written as~\cite{Genon work}
\begin{equation}
\begin{split}
F&=F^{(0)}+F^{(1)}+\ldots =\Phi ^{(0)}\otimes T^{(0)}+\Phi ^{(1)}\otimes T^{(0)}+\Phi ^{(0)}\otimes T^{(1)}+\ldots
\end{split}
\label{eq.2.1}
\end{equation}

At the 1-loop level, we find
\begin{equation}
F^{(1)}=\Phi ^{(1)}\otimes T^{(0)}+\Phi ^{(0)}\otimes T^{(1)}
\label{eq.2.2}
\end{equation}

\begin{figure}
\includegraphics[scale=1]{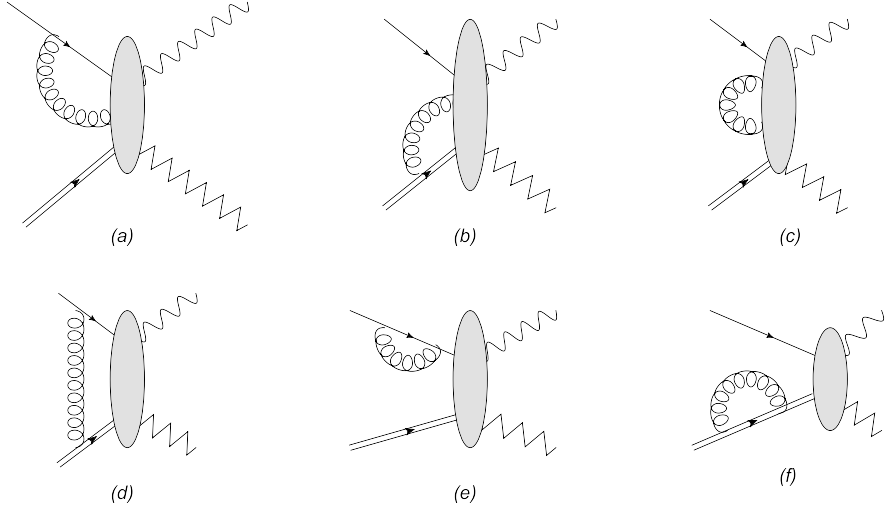}
\caption{\label{nltwist} The 1-loop correction of wave-functions. $\Phi^{(1)} \otimes T_a^{(0)}$ and $\Phi^{(1)} \otimes T_b^{(0)}$ are established in this figure.}
\end{figure}

The 1-loop corrections of wave-function come from the QCD interaction and the Wilson-Line. The later can be written as the path-ordered exponential~\cite{Genon work,Wilson line}
\begin{equation}
\begin{split}
[x,y]&=\mathcal{P}\left[\exp \left(ig_s\int _y^x dz_{\mu} A^{\mu}(z)\right)\right]=\mathcal{P}\left[\sum _{\substack{n}}\frac{(ig_s)^n}{n!}\prod _{\substack{i}}^n\int _y^x dz_{i\mu} A^{\mu}(z_i)\right]
\label{eq.2.3}
\end{split}
\end{equation}
The corrections are shown in Fig.~\ref{nltwist}.

We use $\Phi _q^{(1)}$ to represent the correction with the gluon from the Wilson Line connected to the light quark external leg. So the correction in Fig.~\ref{nltwist}.a can be written as~\cite{Our work}
\begin{equation}
\begin{split}
&\Phi ^{(1)}_{q\;ij}=ig_s^2C_F\int \frac{d^d l}{(2\pi)^d} \frac{1}{l^2} \int _0^1 d\alpha \left(\bar{v}_{\bar{q}}\gamma ^{\rho}\frac{(\slashed {l}+\slashed {p}_{\bar{q}} - m_{\bar{q}})}{(l+p_{\bar{q}})^2-m_{\bar{q}}^2}\right)_j\left(u_Q\right)_i\\
&\times \int \frac{d^dk_q}{(2\pi)^d}\int \frac{d^dk_Q}{(2\pi)^d}  \left(\frac{\partial}{\partial k_Q^{\rho}}-\frac{\partial}{\partial k_q^{\rho}}\right)\left[\delta ^4 (k_q-(p_{\bar{q}}+\alpha l))\delta ^4 (k_Q-(p_Q-\alpha l))\right]\\
\end{split}
\label{eq.2.4}
\end{equation}
where $C_F$ is $(N^2-1)/2N=4/3$ for $SU(N=3)$ gauge group, $i$ and $j$ are the matrix index in spinner space, so
\begin{equation}
\begin{split}
&\Phi_q ^{(1)}\otimes T^{(0)}=ig_s^2C_F\int \frac{d^d l}{(2\pi)^d} \frac{1}{l^2}\bar{v}_{\bar{q}}\gamma ^{\rho}\frac{(\slashed {l}+\slashed {p}_{\bar{q}} - m_{\bar{q}})}{(l+p_{\bar{q}})^2-m_{\bar{q}}^2} \int _0^1 d\alpha\left(\left.\left(\frac{\partial T^{(0)}}{\partial k_q^{\rho}}-\frac{\partial T^{(0)}}{\partial k_Q^{\rho}}\right)\right| _{k_q=k',k_Q=K'}\right)u_Q \\
&k'=p_{\bar{q}}+\alpha l,\;\;K'=p_Q-\alpha l\\
\end{split}
\label{eq.2.4.1}
\end{equation}
Similar to $\Phi _q$, the correction in Fig.~\ref{nltwist}.b is
\begin{equation}
\begin{split}
&\Phi_{Q\;ij} ^{(1)}=-iC_Fg_s^2 \int \frac{d^d l}{(2\pi)^d}\frac{1}{l^2}\int _0^1 d\alpha\left(\bar{v}_{\bar{q}}\right)_j\left(\frac{(\slashed p_Q-\slashed l+m_Q)}{(p_Q-l)^2-m_Q^2}\gamma ^{\rho} u_Q\right)_i\\
&\times \int \frac{d^dk_q}{(2\pi)^d}\int \frac{d^dk_Q}{(2\pi)^d}  \left(\frac{\partial}{\partial k_Q^{\rho}}-\frac{\partial}{\partial k_q^{\rho}}\right)\left[\delta ^4 (k_q-(p_{\bar{q}}+\alpha l))\delta ^4 (k_Q-(p_Q-\alpha l))\right]\\
\end{split}
\label{eq.2.5}
\end{equation}
and
\begin{equation}
\begin{split}
&\Phi_Q ^{(1)}\otimes T^{(0)}=-iC_Fg_s^2 \int \frac{d^d l}{(2\pi)^d}\frac{1}{l^2}\int _0^1 d\alpha
\bar{v}_{\bar{q}}\left(\left.\left(\frac{\partial T^{(0)}}{\partial k_q^{\rho}}-\frac{\partial T^{(0)}}{\partial k_Q^{\rho}}\right)\right| _{k_q=k',k_Q=K'}\right)\\
&\times \frac{(\slashed p_Q-\slashed l+m_Q)}{(p_Q-l)^2-m_Q^2}\gamma ^{\rho} u_Q\\
&k'=p_{\bar{q}}+\alpha l,\;\;K'=p_Q-\alpha l\\
\end{split}
\label{eq.2.5.1}
\end{equation}

We use $\Phi _{\rm Wfc}$ to denote the correction shown in Fig.~\ref{nltwist}.c. We find
\begin{equation}
\begin{split}
&\Phi_{\rm Wfc\;ij}^{(1)}=-i\frac{g_s^2C_F}{2}\int \frac{d^dl}{(2\pi)^d}\frac{1}{l^2}\int _0^1d\alpha\int _0^1 d\beta \left(\bar{v}_{\bar{q}}\right)_j\left(u_Q\right)_i\\
&\times \int \frac{d^dk_q}{(2\pi)^d}\int \frac{d^dk_Q}{(2\pi)^d}  \left(\frac{\partial }{\partial k_{q}}-\frac{\partial }{\partial k_Q}\right)^2 \left[\delta^4(k_q-(p_{\bar{q}}+(\alpha -\beta)l))\delta ^4 (k_Q-(p_Q+(\alpha -\beta)l))\right]\\
\end{split}
\label{eq.2.6}
\end{equation}
and
\begin{equation}
\begin{split}
&\Phi_{\rm Wfc}^{(1)}\otimes T^{(0)}=-i\frac{g_s^2C_F}{2}\int \frac{d^dl}{(2\pi)^d}\int _0^1d\alpha\int _0^1 d\beta \frac{1}{l^2}\bar{v}_{\bar{q}}\left(\left.\left(\frac{\partial }{\partial k_{q}}-\frac{\partial }{\partial k_Q}\right)^2T^{(0)}\right|_{k_q=k',k_Q=K'}\right)u_Q\\
&k'=p_{\bar{q}}+(\alpha -\beta)l, K'=p_Q+(\alpha -\beta)l\\
\end{split}
\label{eq.2.6.1}
\end{equation}

When the wave functions at 1-loop order convolute with the tree level hard function $T^0$, the momenta in $T^0$, $k_Q$ and $k_{\bar{q}}$ will not be on the mass-shell because of the momentum of the gluon that connects with the quark line flowing into it. The $\delta$ functions in Eqs. (\ref{eq.2.4}), (\ref{eq.2.5}) and (\ref{eq.2.6}) clearly show this, where $p_Q$ and $p_{\bar{q}}$ are the momenta of the external on-shell quark. The off-shellness of $k_Q$ and $k_{\bar{q}}$ makes the Eqs. (\ref{eq.1.8}) and (\ref{eq.1.9}) not equivalent. We find that, by using the definition of Eq. (\ref{eq.1.9}) for $T^0$, all the infrared divergence can be absorbed into the wave function. Therefore the definition in Eq. (\ref{eq.1.9}) is used in next sections.

The corrections shown in Figs.~\ref{nltwist}.d, e and f can be denoted as $\Phi _{\rm box}^{(1)}$, $\Phi _{\rm ext q}^{(1)}$ and $\Phi _{\rm ext Q}^{(1)}$. We find that, they have the same forms as the free particle 1-loop QCD box correction and external leg corrections.

\section{\label{sec:loop}1-loop Factorization}

The hard-scattering kernel at 1-loop, denoted as $T^{(1)}$ can be calculated by using the 1-loop QCD corrections of free quarks $F^{(1)}$, and the 1-loop corrections of the wave function $\Phi ^{(1)}\otimes T^{(0)}$ through the relation
\begin{equation}
\begin{split}
&\Phi ^{(0)}\otimes T^{(1)} = F^{(1)}-\Phi ^{(1)}\otimes T^{(0)}\\
\end{split}
\label{eq.3.1}
\end{equation}
For simplicity, we define
\begin{equation}
\begin{split}
&x=m_Q^2,\;\;y=2p_Q\cdot p_{\gamma},\;\;w=2p_Q\cdot p_{\bar{q}},\;\;z=2p_{\gamma}\cdot p_{\bar{q}}\\
\end{split}
\label{eq.3.2}
\end{equation}
For convenience, we will establish our results as $\Phi ^{(0)}\otimes T^{(1)}$ instead of $T^{(1)}$, because the delta function in $\Phi ^{(0)}$ will just replace the transmission momentum $k_Q$ and $k_{\bar{q}}$ in $T^{(1)}$ with the momentums of the quarks $p_Q$ and $p_{\bar{q}}$.

In the calculation, the collinear IR divergences are regulated by assigning a small mass $m_q$ to the light quark. The soft IR divergences will not appear explicitly in the calculations. We use the dimensional regulator (DR) in $D=4-\epsilon$ dimension and $\overline{\rm MS}$ scheme~\cite{msbar} to regulate the ultraviolet (UV) divergences, and $N_{\rm UV}$ is defined as
\begin{equation}
\begin{split}
&N_{\rm UV}=\frac{2}{\epsilon}-\gamma _E+\log (4\pi)
\end{split}
\label{eq.3.3}
\end{equation}
We use the renormalization scale equal to the factorization scale.

\subsection{\label{sec:41}corrections of $F_a^{(1)}$ and $\Phi ^{(1)}\otimes T_a^{(0)}$}

\begin{figure}
\includegraphics[scale=1.0]{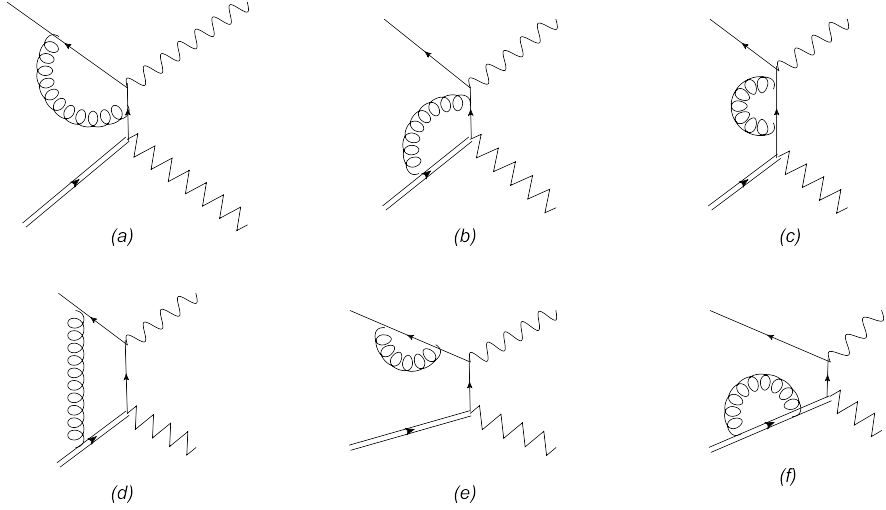}
\caption{\label{1loopa} The 1-loop correction of $F_a^{(1)}$.}
\end{figure}

The Feynman diagrams of $F_a^{(1)}$ and $\Phi ^{(1)}\otimes T_a^{(0)}$ are shown in Fig.~\ref{1loopa} and Fig.~\ref{nltwist}. The diagrams in Figs.~\ref{1loopa}.(a)$\sim$(f) are denoted as $F_a^{(1)\rm EM}$, $F_a^{(1)\rm Weak}$, $F_a^{(1)\rm Wfc}$, $F_a^{(1)\rm box}$, $F_a^{(1)\rm ext\;q}$ and $F_a^{(1)\rm ext\;Q}$ respectively. For simplicity, we define $q=p_{\gamma}-p_{\bar{q}}$.

% \section{\label{sec:level1}4.1.1. EM vertex correction}

The $F_a^{(1)\rm EM}$ is
\begin{equation}
\begin{split}
&F^{(1){\rm EM}}_{a}=ie_qC_Fg_s^2\int \frac{d^d l}{(2\pi)^d}\bar{v}_{\bar{q}}\gamma _{\rho}\frac{(\slashed p_{\bar{q}}+\slashed l - m_q)}{(p_{\bar{q}}+l)^2-m_q^2}\slashed \varepsilon \frac{\slashed l - \slashed q-m_q}{(l-q)^2-m_q^2}\gamma ^{\rho}\frac{\slashed q}{q^2}P_L^{\mu}\frac{1}{l^2}u_Q\\
\end{split}
\end{equation}
and we find
\begin{equation}
\begin{split}
&F_a^{(1)\rm EM}=-ie_qg_s^2C_F\frac{i}{16\pi^2}\bar{v}_{\bar{q}}\frac{\slashed \varepsilon \slashed p_{\gamma}}{q^2} P_L^{\mu}u_Q\times \left(N_{\rm UV}-\log\frac{z}{\mu^2}+2\log\frac{z}{m_q^2}\right)\\
&-ie_qg_s^2C_F\frac{i}{16\pi^2}\bar{v}_{\bar{q}}\frac{\slashed \varepsilon (-\slashed p_{\bar{q}})}{q^2} P_L^{\mu}u_Q\times \left(N_{\rm UV}-\log\frac{z}{\mu^2}+1\right)\\
\end{split}
\label{eq.4.5}
\end{equation}
The corresponding correction of the wave-function to the $F_a^{(1)\rm EM}$ is $\Phi _q^{(1)}\otimes T_a^{(0)}$, which can be written as
\begin{equation}
\begin{split}
&\Phi ^{(1)}_{q}\otimes T^{(0)}_{a}=ie_qC_Fg_s^2\int _0^1d\alpha \int \frac{d^d l}{(2\pi)^d}\frac{1}{l^2}\bar{v}_q \gamma ^{\rho}\frac{\slashed l + \slashed p_q}{(l+p_q)^2-m_q^2} \slashed \varepsilon \\
&\times \left( -\frac{\gamma _{\rho}}{(q-\alpha l)^2 - m_q^2} + \frac{2(q-\alpha l)_{\rho}(\slashed q -\alpha \slashed l)}{((q-\alpha l)^2 - m_q^2)^2}\right)u_Q\\
\end{split}
\end{equation}
Using the method given in Appendix~\ref{sec:ap1}, we find
\begin{equation}
\begin{split}
&\Phi ^{(1)}_{q}\otimes T^{(0)}_{a}=\frac{i}{16\pi^2}\frac{-ie_qC_Fg_s^2\bar{v}\slashed \varepsilon\slashed p_{\gamma}P_L^{\mu}u_Q}{q^2}\left(N_{UV}-\log \frac{z}{\mu^2}+2\log \frac{z}{m_q^2}\right)\\
&+\frac{i}{16\pi^2}\frac{-ie_qC_Fg_s^2\bar{v}\slashed \varepsilon(-\slashed p_{\bar{q}})P_L^{\mu}u_Q}{q^2}\left(N_{UV}-\log \frac{z}{\mu^2}+2\right)\\
\end{split}
\label{eq.4.6}
\end{equation}
So we can obtain
\begin{equation}
\begin{split}
&\Phi ^{(0)}\otimes T_a^{(1)\rm EM}=-ie_qC_Fg_s^2\frac{i}{16\pi^2}\frac{\bar{v}\slashed \varepsilon \slashed p_{\bar{q}} P_L^{\mu}u_Q}{q^2}\\
\end{split}
\label{eq.4.7}
\end{equation}

The correction $F_a^{(1)\rm Wfc}$ is
\begin{equation}
\begin{split}
F_a^{(1)\rm Wfc}=-ie_qC_Fg_s^2\bar{v}_{\bar{q}}\slashed \varepsilon \frac{\slashed q}{q^2}P_L^{\mu}u_Q\left(N_{UV}-\log \frac{z}{\mu} + 1\right)\\
\end{split}
\label{eq.4.8}
\end{equation}
In Ref.~\cite{Genon work,Our work}, the correction of wave-function corresponding to $F_a^{(1)\rm Wfc}$ is found to be $0$. However, using the $T_a^{(0)}$ in Eq.~(\ref{eq.1.9}), we find
\begin{equation}
\begin{split}
&\Phi^{(1)}_{\rm Wfc}\otimes T_a^{(0)}=(D-2)ig_s^2C_F\int\frac{d^D l}{(2\pi)^D}\frac{1}{l^2}\bar{v}_{\bar{q}}\int _0^1d\alpha\int _0^1d\beta \frac{\slashed q - (\alpha -\beta)\slashed l}{(q-(\alpha -\beta)l)^4} u_Q\\
\end{split}
\label{4.9}
\end{equation}
The result is given in Eq.~(\ref{b7})
\begin{equation}
\begin{split}
&\Phi_{\rm Wfc}^{(1)}\otimes T^{(0)}_{a}=-2ig_s^2C_F\bar{v}_{\bar{q}}\slashed \varepsilon \frac{\slashed q}{q^2}P_L^{\mu}u_Q\frac{i}{16\pi^2}\left(N_{\rm UV}-\log \frac{z}{\mu^2}+3\right)\\
\end{split}
\label{4.10}
\end{equation}
correspondingly we find
\begin{equation}
\begin{split}
&\Phi ^{(0)}\otimes T_a^{(1)\rm Wfc}=ie_qC_Fg_s^2\frac{i}{16\pi^2}\frac{\bar{v}\slashed \varepsilon \slashed q P_L^{\mu}u_Q}{q^2}\left(N_{\rm UV}-\log\frac{z}{\mu^2}+5\right)\\
\end{split}
\label{eq.4.11}
\end{equation}

The correction corresponding to $F_a^{(1)\rm Weak}$ is $\Phi _Q^{(1)}\otimes T_a^{(0)}$, which can be written as
\begin{equation}
\begin{split}
&\Phi_Q ^{(1)}\otimes T_a^{(0)}=ie_qC_Fg_s^2\int \frac{d^d l}{(2\pi)^d}\frac{1}{l^2}
\bar{v}_{\bar{q}}\left(\slashed \varepsilon \int_0^1 d\alpha \left(\frac{\gamma _{\rho}}{(\alpha l-q)^2} - \frac{2 (\alpha l -q)^{\rho}\left(\alpha \slashed l -\slashed q \right)}{(\alpha l-q)^4} \right)\right)\\
&\times P_L^{\mu}\frac{(\slashed p_Q-\slashed l+m_Q)}{(p_Q-l)^2-m_Q^2}\gamma ^{\rho} u_Q\\
\end{split}
\label{eq.4.12}
\end{equation}
We calculate Weak vertex correction $F_a^{(1)\rm Weak}$ together with $\Phi _Q^{(1)}\otimes T_a^{(0)}$, and finally we get
\begin{equation}
\begin{split}
&\Phi^{(0)}\otimes T_a^{(1)\rm Weak}=ie_qg_s^2\frac{i}{16\pi^2}\frac{\bar{v}_{\bar{q}}\slashed \varepsilon}{q^2}\left(-\slashed p_Q P_L^{\mu}(4C_{00}-1-B_0(x-y+w-z,x,0))\right.\\
&+(B_0(x,0,x)+B_0(-z,0,0)-2B_0(x-y+w-z,x,0))\slashed q P_L^{\mu}\\
&+(1-q^2C_0+2p_Q^2 C_{11}+4p_Q^2C_{12}-2q^2C_{22}-2p_Q\cdot q C_1-4q^2C_2)\slashed q P_L^{\mu}\\
&+(-4p_Q\cdot q C_{11}-4p_Q\cdot q C_{12}-4q^2C_{12}-2p_Q\cdot q C_1-2q^2C_1-2q^2C_2-q^2C_0)\slashed p_Q P_L^{\mu}\\
&+(-8p_Q^2C_{11}-8p_Q^2C_{12}+12q^2C_{22}-4p_Q^2C_1+16q^2C_2)p_Q^{\mu}(1-\gamma _5)\\
&+(4q^2C_0+16p_Q\cdot q C_{11}+16p_Q\cdot q C_{12}+12q^2C_{12}+8p_Q\cdot qC_1+8q^2C_1)p_Q^{\mu}(1-\gamma _5)\\
&+(-4C_{11}-8C_{22}-12C_{12}-8C_2-4C_1)q^{\mu}\slashed q (1+\gamma _5)m_Q\\
&\left.+(2q^2 \slashed p_Q P_L^{\mu} -2q\cdot p_Q \slashed q)\left(2a_1-C_2-2C_{22}\right)\right)u_Q\\
\end{split}
\label{eq.4.13}
\end{equation}
where $B_0$, $C_i$ and $C_{ij}$ are scalar Passarino-Veltman functions(Pa-Ve function)~\cite{trangle calc} and defined as
\begin{equation}
\begin{split}
&B_0(r_{10}^2;m_0^2,m_1^2)=\frac{(2\pi \mu)^\epsilon}{i\pi^2}\int d^dk \prod _{\substack{i=0}}^1\frac{1}{(k+r_i)^2-m_i^2}\\
\end{split}
\label{eq.4.14}
\end{equation}
\begin{equation}
\begin{split}
&C_0(r_{10}^2,r_{12}^2,r_{20}^2;m_0^2,m_1^2,m_2^2)=\frac{(2\pi \mu)^\epsilon}{i\pi^2}\int d^dk \prod _{\substack{i=0}}^2\frac{1}{(k+r_i)^2-m_i^2}
\end{split}
\label{eq.4.15}
\end{equation}
and $C_1$, $C_2$ and $C_{ij}$ are defined in $C^{\mu}$ and $C^{\mu\nu}$ with
\begin{equation}
\begin{split}
&C^{\mu}=r_1^{\mu}C_1+r_2C_2\\
&C^{\mu\nu}=g^{\mu\nu}C_{00}+\sum _{\substack{i,j=1}}^2r_i^{\mu}r_j^{\nu}C_{ij}
\end{split}
\label{eq.4.16}
\end{equation}
with
\begin{equation}
\begin{split}
&C^{\mu}(r_{10}^2,r_{12}^2,r_{20}^2;m_0^2,m_1^2,m_2^2)=\frac{(2\pi \mu)^\epsilon}{i\pi^2}\int d^dk k^{\mu}\prod _{\substack{i=0}}^2\frac{1}{(k+r_i)^2-m_i^2}\\
&C^{\mu\nu}(r_{10}^2,r_{12}^2,r_{20}^2;m_0^2,m_1^2,m_2^2)=\frac{(2\pi \mu)^\epsilon}{i\pi^2}\int d^dk k^{\mu}k^{\nu}\prod _{\substack{i=0}}^2\frac{1}{(k+r_i)^2-m_i^2}
\end{split}
\label{eq.4.17}
\end{equation}
In the expression above, the function $C$'s are short for $C(-z,x-y+w-z,x;0,0,x)$'s.

And $a_1$ in Eq.~(\ref{eq.4.13}) is defined as:
\begin{equation}
\begin{split}
&a_1=m_Q^2\int _0^1 dx\int _0^{1-x}dy\int _0^1 d\alpha \frac{\alpha ^2 y^2}{(\alpha ^2 y^2 m_Q^2+2\alpha x y p_Q\cdot q+x(1-x)(-q^2))^2}
\end{split}
\label{eq.4.18}
\end{equation}
There is no IR Div caused by gluon in those integrals. For convenience, we will give the result of $C_0$, $C_{ij}$, $B_0$ and $a_1$ up to the sub-leading order after expansion in the next section.

When taking the renormalization scale equal to the factorization scale, the corrections of the external legs and the box diagram are equal to the relevant corrections to the wave-function, then we can obtain
\begin{equation}
\begin{split}
&T_a^{(1)\rm ext\;q}=T_a^{(1)\rm ext\;Q}=T_a^{(1)\rm box}=0\\
\end{split}
\label{eq.4.19}
\end{equation}

\subsection{\label{sec:42}corrections of $F_b^{(1)}$ and $\Phi ^{(1)}\otimes T_b^{(0)}$}

\begin{figure}
\includegraphics[scale=1.0]{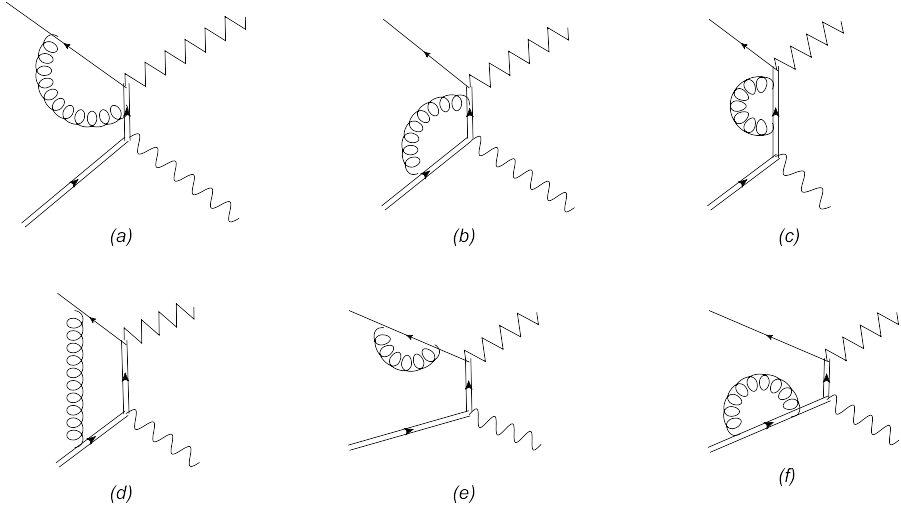}
\caption{\label{1loopb} The 1-loop correction of $F_b^{(1)}$.}
\end{figure}

The Feynman diagrams of $F_b^{(1)}$ and $\Phi ^{(1)}\otimes T_b^{(0)}$ are shown in Fig.~\ref{1loopb} and Fig.~\ref{nltwist}. The diagrams of Figs.~\ref{1loopb}.(a)$\sim$(f) are denoted as $F_b^{(1)\rm Weak}$, $F_b^{(1)\rm EM}$, $F_b^{(1)\rm Wfc}$, $F_b^{(1)\rm box}$, $F_b^{(1)\rm ext\;q}$ and $F_b^{(1)\rm ext\;Q}$ respectively. For simplicity, we define $Q=p_Q-p_{\gamma}$.

There are collinear IR divergences in $F_b^{(1)\rm Weak}$ in higher orders of $O(\Lambda _{\rm QCD}\left./\right.m_Q)$ corrections. When $E_{\gamma}\sim m_Q$ or $E_{\gamma}\sim \Lambda _{\rm QCD}^2\left./\right.m_Q$, $F_b^{(1)\rm Weak}$ is at the order of $O(\left.\Lambda _{\rm QCD}/m_Q\right)^2$, while $E_{\gamma}\sim \Lambda _{\rm QCD}$, $F_b^{(1)\rm Weak}$ is at order $O(\left.\Lambda _{\rm QCD}/m_Q\right)$. The relevant correction of the wave-function is $\Phi _q^{(1)}\otimes T_b^{(0)}$, which can be written as
\begin{equation}
\begin{split}
&\Phi_q^{(1)}\otimes T^{(0)}_{b}=-ie_Qg_s^2\int\frac{d^d l}{(2\pi)^d}\frac{1}{l^2}\bar{v}_{\bar{q}}\gamma ^{\rho}\frac{\slashed l + \slashed p_{\bar{q}}}{(l+p_{\bar{q}})^2-m_q^2}P_L^{\mu}\\
&\times \int_0^1 d\alpha \left(\frac{\gamma _{\rho}}{(Q-\alpha l)^2-m_Q^2} - \frac{2 (Q-\alpha l )^{\rho}\left(\slashed Q -\alpha \slashed l  +m_Q\right)}{((Q-\alpha l)^2-m_Q^2)^2} \right)\slashed \varepsilon u_Q\\
\end{split}
\label{eq.4.20}
\end{equation}
The hard scattering kernel $\Phi ^{(0)}\otimes T_b^{(1)\rm Weak}=F_b^{(1)\rm Weak} - \Phi_q^{(1)}\otimes T^{(0)}_{b}$ is given in Eq.~(\ref{eq.a32}). We find the collinear IR divergences are canceled to all orders, and the result is
\begin{equation}
\begin{split}
&\Phi ^{(0)}\otimes T_b^{(1)\rm Weak}=ie_Qg_s^2C_F\bar{v}_{\bar{q}}\\
&\times \left(\left(\frac{y-w+z}{w-z}\left(B_0(x-y,0,x)-B_0(x-y+w-z,0,x)\right)+2\right) P_L^{\mu}\frac{\slashed Q+m_Q}{Q^2-m_Q^2}\right.\\
&\left.+2\slashed Q\gamma^{\mu}(1+\gamma _5)C_2-4p_{\bar{q}}^{\mu}(1+\gamma _5)\frac{Q^2+\slashed Q m_Q}{Q^2-m_Q^2}C_2-4(p_{\bar{q}}^{\mu}C_{12}-Q^{\mu}C_{22})(1+\gamma _5)\frac{\slashed Q m_Q+Q^2}{Q^2-m_Q^2}\right.\\
&\left.+f_1\left(\left(\frac{w-z}{y}+\frac{w-z}{x-y}\right)\slashed Q + \frac{w-z}{y} m_Q\right)+f_2+f_3\right)\slashed \varepsilon u_Q\\
\end{split}
\label{eq.4.21}
\end{equation}
where the $C_i$'s and $C_{ij}$'s are short for the Pa-Ve functions $C_i(0,x-y+w-z,x; 0,0,x)$ and $C_{ij}(0,x-y+w-z,x; 0,0,x)$. And $f_1$, $f_2$, $f_3$ are defined in Eqs.~(\ref{eq.f1}), (\ref{eq.a6}) and (\ref{eq.a27}).

The result of $F_b^{(1)\rm Wfc}$ is
\begin{equation}
\begin{split}
&F_{b}^{(1)\rm Wfc}=-ie_QC_Fg_s^2\bar{v}_{\bar{q}}\frac{i}{16\pi^2}P_L^{\mu}\left(\frac{(2x-y)\slashed Q+(2x-2y)m_Q}{y^2}\left(N_{\rm UV}-\log \frac{x}{\mu^2}+2\right.\right.\\
&\left.\left.-\frac{y(2x-y)}{(x-y)^2}\log\frac{x}{y}+\frac{x}{x-y}\right)\right.\\
&\left.+\frac{\slashed Q + m_Q}{y}\left(2N_{\rm UV}-2\log \frac{x}{\mu^2}+2-2\frac{y(2x-y)}{(x-y)^2}\log\frac{x}{y}+2\frac{x}{x-y}\right)\right.\\
&\left.+\frac{2x\slashed Q+(2x-y)m_Q}{y^2}\left(2N_{\rm UV}-2\log \frac{y}{\mu^2}+1-2\frac{x(x-2y)}{(x-y)^2}\log\frac{x}{y}-2\frac{y}{x-y}\right)\right)\slashed \varepsilon u_Q\\
\end{split}
\label{eq.4.22}
\end{equation}
Notice that, when $E_{\gamma}$ is small, $F_b^{(0)}\sim O(1/\Lambda _{\rm QCD})$ is at sub-leading order, however, $F_b^{(1)\rm Wfc}\sim O\left( m_Q/ \Lambda _{\rm QCD}^2\right)$ is larger then leading order. This behavior is new for the soft photon case. We will discuss it in the next section.

The corresponding correction of the wave function is $\Phi _{\rm Wfc} ^{(1)}\otimes T_b^{(0)}$, which can be written as
\begin{equation}
\begin{split}
&\Phi^{(1)}_{{\rm Wfc}}\otimes T^{(0)}_{b}=-i2g_s^2e_QC_f\int\frac{d^D l}{(2\pi)^D}\frac{1}{l^2}\bar{v}_{\bar{q}}P_L^{\mu}\int _0^1d\alpha\int _0^1d\beta \left(\frac{2m_Q^2(\slashed Q +(\alpha -\beta)\slashed l+m_Q)}{((Q+(\alpha -\beta)l)^2-m_Q^2)^3}\right.\\
&\left.-(\frac{D-2}{2})\frac{(\slashed Q +(\alpha -\beta)\slashed l)}{((Q+(\alpha -\beta)l)^2-m_Q^2)^2}-(\frac{D-4}{2})\frac{m_Q}{((Q+(\alpha -\beta)l)^2-m_Q^2)^2}\right)\slashed \varepsilon u_Q\\
\end{split}
\label{eq.4.23}
\end{equation}
the result is
\begin{equation}
\begin{split}
&\Phi^{(1)}_{{\rm Wfc}}\otimes T^{(0)}_{b}=-i2g_s^2C_F\frac{i}{16\pi^2}\bar{v}_{\bar{q}}P_L^{\mu}\left(\frac{m_Q}{y}\left(\frac{x}{x-y}\log\frac{x}{y}+\frac{y}{x-y}-\frac{x^2}{(x-y)^2}\log\frac{x}{y}\right)\right.\\
&\left.+\frac{\slashed Q+m_Q}{y}\left(-N_{\rm UV}+\log\frac{y}{\mu^2}-\frac{8x}{x-y}\log\frac{x}{y}+\frac{y}{x-y}-\frac{x^2 \log \frac{x}{y}}{(x-y)^2}-2\right)\right)\slashed \varepsilon u_Q\\
\end{split}
\label{eq.4.24}
\end{equation}
Then the kernel can be obtained by using Eq.~(\ref{eq.3.1}).

The result of $F_b^{(1)\rm EM}$ is obtained to be
\begin{equation}
\begin{split}
&F_{b}^{(1)\rm EM}=ie_QC_fg_s^2\frac{i}{16\pi^2}\bar{v}_{\bar{q}}P_L^{\mu}\frac{\slashed Q + m_Q}{y}\left(\left( N_{\rm UV} - 8 -\log\frac{x}{\mu^2}-\frac{10x}{x-y} \log\frac{x}{y} + \frac{7y}{x-y}\log\frac{x}{y}\right.\right.\\
&\left.\left.+\frac{5 x - y}{y}\left(-2{\rm Li}_2(1-\frac{y}{x})+\frac{\pi^2}{3}\right)\right)\slashed \varepsilon\right.\\
&\left.+\left(  \frac{2 x + y}{2 (x - y) y}\log\frac{y}{x} + \frac{1}{y^2}\left(\frac{\pi^2}{6}x - y - x {\rm Li}_2(1-\frac{y}{x})\right)\right)4p_Q\cdot \varepsilon \slashed p_{\gamma} +\frac{1}{x-y}\log\frac{x}{y}2m_Q\slashed p_{\gamma}\slashed \varepsilon\right.\\
&\left.+\left(\frac{4 x - 5 y}{2 (x - y)^2}\log\frac{y}{x}-\frac{1}{2(x-y)} + \frac{1}{y} \left(\frac{\pi^2}{2}-3{\rm Li}_2(1-\frac{y}{x})\right)\right)4m_Q p_Q\cdot \varepsilon\right)u_Q\\
\end{split}
\label{eq.4.25}
\end{equation}
The corresponding correction of wave function is
\begin{equation}
\begin{split}
&\Phi_Q ^{(1)}\otimes T_b^{(0)}=ie_QC_Fg_s^2\int \frac{d^d l}{(2\pi)^d}\frac{1}{l^2}
\bar{v}_{\bar{q}}P_L^{\mu}\\
&\times \int_0^1 d\alpha \left(\frac{\gamma _{\rho}}{(Q-\alpha l)^2-m_Q^2} - \frac{2 (Q-\alpha l )^{\rho}\left(\slashed Q -\alpha \slashed l+m_Q\right)}{((Q-\alpha l)^2-m_Q^2)^2} \right)\slashed \varepsilon \frac{(\slashed p_Q-\slashed l+m_Q)}{(p_Q-l)^2-m_Q^2}\gamma ^{\rho} u_Q\\
\end{split}
\label{eq.4.26}
\end{equation}
The result of the momentum-integration is
\begin{equation}
\begin{split}
&\Phi_Q ^{(1)}\otimes T_b^{(0)}=ie_QC_Fg_s^2\bar{v}_{\bar{q}}P_L^{\mu}\frac{i}{16\pi^2}\left(2\left(\slashed p_Q +\frac{2x-y}{y}(\slashed Q +m_Q)\right)(C_0+d_1)+2(-(2x-y)\slashed p_{\gamma}+y \slashed p_Q) d_2\right.\\
&\left.-d_3\slashed p_{\gamma} + d_4 m_Q + d_5(\slashed p_Q+m_Q)-(\frac{\slashed Q +m_Q}{-y}B_0(x,x,0)-(\slashed Q + m_Q)C_0+\gamma_{\mu} C^{\mu}+d_6)\right)\slashed \varepsilon u_Q\\
\end{split}
\label{eq.4.27}
\end{equation}
where the $C_0$ and $C_{\mu}$ are short for Pa-Ve functions $C_0(x-y,0,x; 0,x,x)$ and $C_{\mu}(x-y,0,x; 0,x,x)$, and
\begin{equation}
\begin{split}
&d_1=-\left(\frac{i \left(2 \rm{Li}_2(\frac{2 y}{y+i \sqrt{(4 x-y) y}})-2 \rm{Li}_2(\frac{2 i y}{i y+\sqrt{(4 x-y) y}})\right)}{\sqrt{y (4 x-y)}}\right.\\
&\left.-\frac{\rm{Li}_2(\frac{y}{x})+\log \frac{x-y}{x} \log \frac{y}{x}}{y}-\frac{2 \log \frac{x}{y} \cot ^{-1}(\frac{y-2 x}{\sqrt{y (4 x-y)}})}{\sqrt{y (4 x-y)}}\right)
\end{split}
\label{eq.4.28}
\end{equation}
\begin{equation}
\begin{split}
&d_2 =\frac{x}{y^3} \left(2 \rm{Li}_3(1-\frac{y}{x})-3 \rm{Li}_2(1-\frac{y}{x})-2 \rm{Li}_3(\frac{y}{x})+2 \rm{Li}_2(\frac{y}{x}) \log \frac{y}{x}-\log \frac{x}{x-y} \log ^2\frac{x}{y}-2 \zeta (3)+\frac{\pi ^2}{2}\right)\\
&+\frac{1}{y^2}\left(\frac{y-2 x}{x-y} \rm{Li}_2(1-\frac{x}{y})+\log \frac{y}{x}-1\right)\\
\end{split}
\label{eq.4.29}
\end{equation}
\begin{equation}
\begin{split}
&d_3=\frac{2}{y}\left(\left(\frac{x}{y} \left(2 \rm{Li}_2(1-\frac{y}{x})-\frac{\pi ^2}{3}\right)+\log \frac{x}{y}+2\right)\right)\\
&+\frac{x}{(x-y)^2} \left(2 \rm{Li}_2(1-\frac{y}{x})+4 \rm{Li}_2(1-\frac{x}{y})+\log \frac{x}{y} \left(\frac{2 x}{y}+3 \log \frac{x}{y}\right)\right)\\
\end{split}
\label{eq.4.30}
\end{equation}
\begin{equation}
\begin{split}
&d_4=\frac{2}{y}\left(\frac{\pi ^2}{6}-\rm {Li}_2(1-\frac{y}{x})\right)+\frac{2y}{(x-y)^2}\left(-\rm {Li}_2(1-\frac{y}{x})-2\rm {Li}_2(1-\frac{x}{y})-\log \frac{x}{y}(\frac{3}{2} \log \frac{x}{y}+\frac{x}{y})\right)\\
\end{split}
\label{eq.4.31}
\end{equation}
\begin{equation}
\begin{split}
&d_5=\frac{2x}{(x-y)^2}\left(\rm{Li}_2(1-\frac{y}{x})+2\rm{Li}_2(1-\frac{x}{y})+\log \frac{x}{y}(\frac{3}{2}\log \frac{x}{y}+1)\right)\\
\end{split}
\label{eq.4.32}
\end{equation}
and
\begin{equation}
\begin{split}
&d_6=\frac{1}{y^2}\left(\left( (2 x-y)(\frac{\pi ^2}{3} -2\rm{Li}_2(1-\frac{y}{x}))-6y-4 y \log \frac{x}{y}+\frac{6xy}{(x-y)^2}\left(y \log \frac{x}{y}- (x-y)\right)\right)\slashed p_{\gamma}\right.\\
&\left.+\left(\frac{\frac{\pi ^2}{3}-2 \rm {Li}_2(1-\frac{y}{x})}{y}-\frac{3 \log \frac{x}{y}+4}{x-y}+\frac{4 y \log \frac{x}{y}}{(x-y)^2}\right)m_Q\right.\\
&\left.-\frac{4 (y \log \frac{x}{y}-x+y)}{y (x-y)^2} m_Q\slashed p_{\gamma}(\slashed p_Q+m_Q)-\frac{\log\frac{x}{y}}{x-y}(\slashed p_Q+m_Q)\right)\\
\end{split}
\label{eq.4.33}
\end{equation}
The kernel can be obtained by using Eq.~(\ref{eq.3.1}).

The case of $T_a$ is similar, we find
\begin{equation}
\begin{split}
&T_b^{(1)\rm ext\;q}=T_b^{(1)\rm ext\;Q}=T_b^{(1)\rm box}=0\\
\end{split}
\label{eq.4.34}
\end{equation}

\subsection{\label{sec:43}corrections of $F_c^{(1)}$ and $\Phi ^{(1)}\otimes T_c^{(0)}$ and other corrections}

\begin{figure}
\includegraphics[scale=0.8]{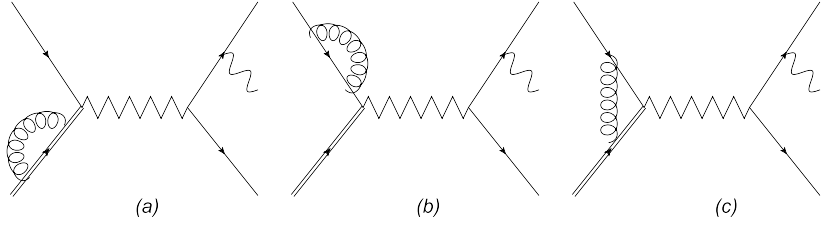}
\caption{\label{1loopc} The 1-loop correction of $F_c^{(1)}$.}
\end{figure}

\begin{figure}
\includegraphics[scale=1.0]{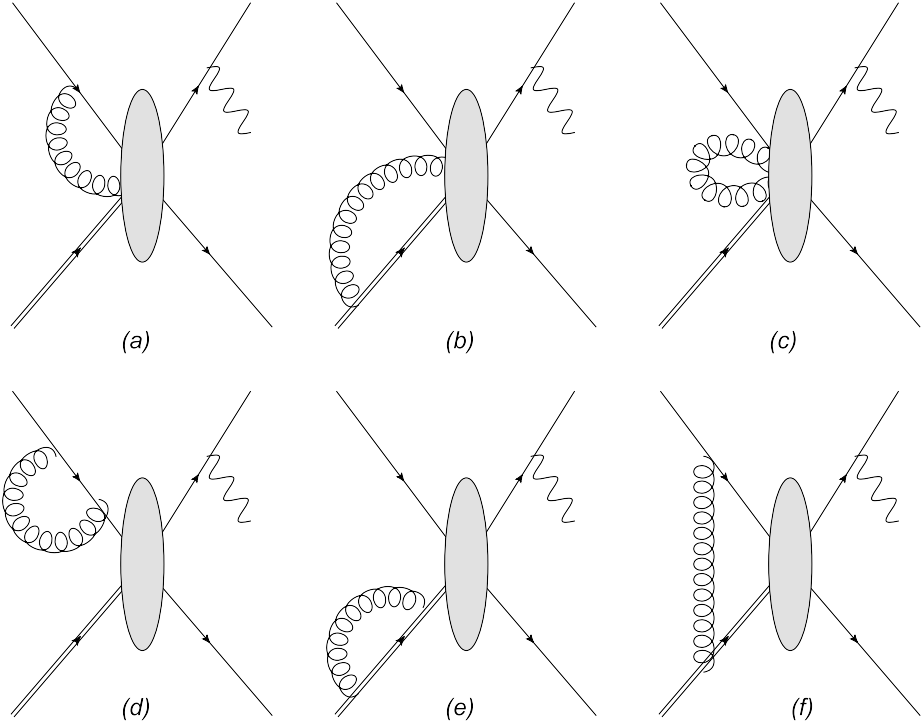}
\caption{\label{twistc} The 1-loop correction of $\Phi ^{(1)}\otimes T_c^{(0)}$.}
\end{figure}

The 1-loop corrections of $F_c^{(0)}$ are show in Fig.~\ref{1loopc}. We denote Figs.~\ref{1loopc}~(a), (b) and (c) as $F_c^{(1)\rm ext\;Q}$, $F_c^{(1)\rm ext\;q}$ and $F_c^{(1)\rm triangle}$ respectively. And the corrections of the wave-function are shown in Fig.~\ref{twistc}. Similar as in Sec.~\ref{sec:wavefunction}, we denote QCD 1-loop corrections to the wave-function in Figs.~\ref{twistc}~(a)$\sim$(f) as $\Phi _q^{(1)}\otimes T_c^{(0)}$, $\Phi _Q^{(1)}\otimes T_c^{(0)}$, $\Phi _{\rm Wfc}^{(1)}\otimes T_c^{(0)}$, $\Phi _{\rm ext\;q}^{(1)}\otimes T_c^{(0)}$, $\Phi _{\rm ext\;Q}^{(1)}\otimes T_c^{(0)}$ and $\Phi _{\rm triangle}^{(1)}\otimes T_c^{(0)}$ respectively.

After calculating $\Phi _{\rm ext\;q}^{(1)}\otimes T_c^{(0)}$, $\Phi _{\rm ext\;Q}^{(1)}\otimes T_c^{(0)}$ and $\Phi _{\rm triangle}^{(1)}\otimes T_c^{(0)}$, the hard kernels we get are similar to Eqs.~(\ref{eq.4.19}) and Eq.~(\ref{eq.4.34}). The results are
\begin{equation}
\begin{split}
T_c^{(1){\rm ext\;q}}=T_c^{(1){\rm ext\;Q}}=T_c^{(1){\rm triangle}}=0
\end{split}
\label{eq.4.35}
\end{equation}

$\Phi _q^{(1)}\otimes T_c^{(0)}$, $\Phi _Q^{(1)}\otimes T_c^{(0)}$, $\Phi _{\rm Wfc}^{(1)}\otimes T_c^{(0)}$ do not have correspondent 1-loop QCD corrections. And because the momentum of quarks appear together as $k_Q+k_{\bar{q}}$ in $T_c^{(0)}$, we find
\begin{equation}
\begin{split}
&\left(\frac{\partial }{\partial k_Q}-\frac{\partial }{\partial k_{\bar{q}}}\right)T_c^{(0)}=0
\end{split}
\label{eq.4.36}
\end{equation}
so we find
\begin{equation}
\begin{split}
&T_c^{(1){\rm q}}=T_c^{(1){\rm Q}}=T_c^{(1){\rm Wfc}}=0\\
\end{split}
\label{eq.4.37}
\end{equation}

\subsection{\label{sec:44}1-loop result summery}

With Eqs.~(\ref{eq.4.7}), (\ref{eq.4.11}), (\ref{eq.4.13}), (\ref{eq.4.19}), (\ref{eq.4.21}), (\ref{eq.4.22}), (\ref{eq.4.24}), (\ref{eq.4.25}), (\ref{eq.4.27}), (\ref{eq.4.34}), (\ref{eq.4.35}) and (\ref{eq.4.37}), we can obtain the 1-loop order hard-scattering kernel. We find that, $T^{(1)}$ is IR finite, and the factorization is proved explicitly to any orders of $O(\Lambda _{\rm QCD}\left./\right.m_Q)$.

A remark about the factorization is given here. There may be hard dynamics contained in the wave function defined in Eq. (\ref{eq.1.1}) when the one-loop QCD correction to the wave function is considered. Hard scale $m_Q$ can appear in the one-loop QCD corrections to the wave function when the gluon is attached to the heavy quark line. Such dynamics is also subtracted from the hard amplitude by using Eq. (\ref{eq.3.1}). Then the one-loop corrected hard amplitude $T^{(1)}$ is not only free from the infrared divergence, but also free from the hard dynamics corresponding to external state. Although there may contain some hard dynamics in the wave function, the wave function is process-independent, i.e., it is universal. In treating meson decays, the wave function of meson should be obtained from nonperturbative method.

Also, we didn't assume the energy of photon $E_{\gamma}$ to be large. So we can extend the valid region of factorization in Refs.~\cite{Genon work,Our work} to $E_{\gamma}\sim \left. \Lambda^2 _{\rm QCD}/ \right. m_Q$. In such region, we can investigate the soft photon IR Div in radiative decay.

In the next section, we will investigate both the hard photon region and the soft photon region, and the result will be established up to the sub-leading order.

\section{\label{sec:resum}expansion and large logarithms}

There are large logarithms and double large logarithms in $\Phi ^{(0)}\otimes T^{(1)}$. Those large logarithms may break the $\alpha _s$ perturbation. To obtain the reliable result of the decay amplitude, those large logarithms should be resummed~\cite{large logarithm and resume}. We concentrate on these large logarithms at the leading order of $\Lambda _{\rm QCD}\left./m_Q\right.$, because those higher order terms are not large and will not affect the $\alpha _s$ expansion.

We concentrate on the contribution of the hard scattering kernel to the amplitude. The amplitude can be obtained by replacing the wave-function with the one obtained in Ref.~\cite{Teacher Yang wave function}.
\begin{equation}
\begin{split}
&\Phi ^{(0)}(k_q, k_Q)=\frac{1}{\sqrt{3}}\int d^3 k \Psi (k) \frac{1}{\sqrt{2}}\sum _{\substack{i}}\left(b_Q^{i+}(\vec{k},\uparrow)d_q^{i+}(-\vec{k},\downarrow)-b_Q^{i+}(\vec{k},\downarrow)d_q^{i+}(-\vec{k},\uparrow)\right)\mid 0\rangle \\
&\times \delta ^3(\vec{k}_{\bar{q}} + \vec{k})\delta ^3(\vec{k}_Q - \vec{k})\delta (k_{\bar{q}0}-\sqrt{k^2+m_q^2})\delta (k_{Q0}-\sqrt{k^2+m_Q^2})
\end{split}
\label{eq.5.1}
\end{equation}
with
\begin{equation}
\begin{split}
&\Psi (\vec{k})=4\pi \sqrt{m_P \lambda _P ^3} e^{-\lambda _P |\vec{k}|}
\end{split}
\label{eq.5.2}
\end{equation}

The matrix element can be written as
\begin{equation}
\begin{split}
&F^{\mu}(\mu)=\frac{1}{(2\pi)^3}\frac{3}{\sqrt{6}}\int d^3k \int d^4 k_Q \int d^4 k_{\bar{q}} \Psi (k)  Tr\left[M\cdot \left(T^{(0)\mu}(k_{\bar{q}}, k_Q)+T^{(1)\mu}(k_{\bar{q}}, k_Q)\right)\right]\\
&\times \delta ^3(\vec{k}_{\bar{q}} + \vec{k})\delta ^3(\vec{k}_Q - \vec{k})\delta (k_{\bar{q}0}-\sqrt{k^2+m_q^2})\delta (k_{Q0}-\sqrt{k^2+m_Q^2})
\end{split}
\label{eq.5.3}
\end{equation}
with $M=u_Q(\vec{k}, \uparrow)\bar{v}_q(-\vec{k},\downarrow)-u_Q(\vec{k}, \downarrow)\bar{v}_q(-\vec{k},\uparrow)$ is the notation for Dirac spinner of the meson. The functions $T^{(0)}$ and $T^{(1)}$ are coefficients which are scalars in spinner space multiplied by some simple Dirac structures. So it is natural to write them in the form of products of coefficients and operators. After convolution, $F^{\mu}$ can be written as
\begin{equation}
\begin{split}
&F^{\mu}(\mu)=\sum _{\substack{n}}\frac{1}{(2\pi)^3}\frac{3}{\sqrt{6}}\int d^3k \Psi (k) Tr\left[M\cdot \left(C_n(p_Q, p_{\bar{q}}, \mu)O_n(p_Q, p_{\bar{q}}\right))\right]
\end{split}
\label{eq.5.4}
\end{equation}
with $p_q=(\sqrt{m_q^2 + k^2}, -\vec{k})$ and $p_Q=(\sqrt{m_Q^2 +k^2}, \vec{k})$ denote the on-shell momenta of the light anti-quark and the heavy quark in the bound state.

After the convolution, the Dirac structures can be simplified. The matrix element can be decomposed as \cite{Genon work,bsw1}
\begin{equation}
\begin{split}
&\langle \gamma\mid \bar{q} \Gamma ^{\mu} Q\mid P\rangle=\epsilon _{\mu\nu\rho\sigma}\varepsilon ^{\nu}p_P^{\rho}p_{\gamma}^{\sigma}F_V+i\left(\varepsilon ^{\mu}p_P\cdot p_{\gamma}-p_{\gamma}^{\mu}\varepsilon \cdot p_P\right)F_A\\
\end{split}
\label{eq.5.5}
\end{equation}
To calculate the form factors $F_A$ and $F_V$ in Eq.~(\ref{eq.5.5}), we introduce $6$ operators
\begin{equation}
\begin{split}
&O_1=e_q\slashed \varepsilon \frac{ \slashed p_{\gamma}}{(p_{\gamma}-p_{\bar{q}})^2}P_L^{\mu},\;\;O_2=-e_q\slashed \varepsilon\frac{ \slashed p_{\bar{q}}}{(p_{\gamma}-p_{\bar{q}})^2}P_L^{\mu},\\
&O_3=e_QP_L^{\mu}\frac{\slashed p_Q+m_Q}{(p_Q-p_{\gamma})^2-m_Q^2}\slashed \varepsilon,\;\;O_4=e_QP_L^{\mu}\frac{-\slashed p_{\gamma}}{(p_Q-p_{\gamma})^2-m_Q^2}\slashed \varepsilon\\
&O_5=e_QP_L^{\mu}\frac{m_Q}{(p_Q-p_{\gamma})^2-m_Q^2}\slashed \varepsilon\\
&O_6=-e\frac{-\slashed p_{\gamma }\slashed \varepsilon _{\gamma }^*}{2p_{\gamma}\cdot p_l}P_L^{\mu }+eP_L^{\mu}\left(-\frac{\varepsilon \cdot p_l}{p_{\gamma }\cdot p_l}\right)\\
\end{split}
\label{eq.5.6}
\end{equation}

Then we can concentrate on those $C_nO_n$ terms. The large logarithms are related to the power counting and expansion, and the power counting and expansion are different in soft and hard photon region, so we discuss them separately.

\subsection{\label{sec:51}the hard photon region}

When $E_{\gamma} \sim m_Q$, keeping only the leading order and the sub-leading order terms, we find
\begin{equation}
\begin{split}
&C_1^0=1-\frac{\alpha _s C_F}{4\pi}\left(6+\frac{\pi^2}{3}+\frac{y-2x}{x}\log\frac{x}{y}-2{\rm Li}_2\frac{y}{x} - \log^2\frac{x}{z}+\log\frac{x}{z}\right.\\
&\left.+2\log\frac{x}{y}\log\frac{x-y}{z}-\log\frac{z}{\mu^2}\right)\\
\end{split}
\label{eq.5.7}
\end{equation}
\begin{equation}
\begin{split}
&C_1^1=-\frac{\alpha_s C_F}{4\pi}\left(\frac{2}{y^2}\left(\text{Li}_2(\frac{y}{x}) (-2 w y+2 x z \log \frac{y}{x}+8 x z+y z)-4 x z \text{Li}_3(\frac{y}{x})\right.\right.\\
&\left.\left.+(4 x z-2 w y)\log \frac{x}{y} (\log \frac{x}{z}-2 \log \frac{x-y}{y})-2 x z \zeta (3)\right)+\frac{2 w y}{(x-y)^2}\left(5 \log \frac{y}{x}+6 \log \frac{y}{z}\right)\right.\\
&\left.-\frac{2 w x}{y (x-y)}\left(-\frac{(2 x-y) (x z-w y)}{2 w x^2}-2 \log \frac{x^2 z}{y^3}+\log \frac{x}{y} \log \frac{z (x-y)^2}{x y^2}-\frac{6 x \log \frac{y^2}{x z}}{x-y}-\frac{5 \pi ^2}{6}+4\right)\right.\\
&\left.-\frac{1}{x (x-y)}\left(w y \left(1-\frac{x \log \frac{x^7 z^4}{y^{11}}}{x-y}\right)+wx\left(\frac{x \left(4 \log \frac{x^2 z}{y^3}-26 \log \frac{x}{y}-24 \log \frac{z}{y}\right)}{x-y}\right.\right.\right.\\
&\left.\left.\left.+2 \log \frac{x}{y} \left(2 \log \frac{y}{x-y}+\log \frac{x}{z}\right)-2 \log \frac{x}{y}+\frac{5 \pi ^2}{3}-7\right)\right)\right.\\
&\left.-\frac{2 x^2 z}{y^2 (x-y)}\left(\frac{1}{3} \pi ^2 \log \frac{x y^2}{z^3}-\log \frac{y}{z} \left(\log ^2\frac{x}{z}-2 \log \frac{x z}{y^2}\right)+\frac{(6 x-8 y) \log \frac{y^2}{x z}}{x-y}+\frac{1}{3} \log ^3\frac{x}{z}+\frac{4}{3} \log ^3\frac{y}{z}\right.\right.\\
&\left.\left.+\frac{7 \pi ^2}{3}-6-\frac{y}{x-y}\left(-11 \log \frac{x}{y}-10 \log \frac{z}{y}+\frac{10 \pi ^2}{3}-8\right)\right)\right.\\
&\left.-\frac{z}{x-y}\left(2 \log \frac{y}{x-y} \log \frac{x}{y}+\log ^2\frac{x}{z}-2 \log ^2\frac{y}{z}+5 \log \frac{z}{y}-\frac{4 \pi ^2}{3}+2\right.\right.\\
&\left.\left.-\frac{2 x \left(9 \log \frac{x}{y}+10 \log \frac{z}{y}-\frac{10 \pi ^2}{3}+8\right)}{x-y}\right)\right.\\
&\left.+\frac{x z}{y (x-y)}\left(\frac{x \log \frac{x^3 z^4}{y^7}}{x-y}-2 \log \frac{x}{y} \log \frac{z^2 (x-y)}{y^3}+\frac{2 \pi ^2}{3} \log \frac{x y^2}{z^3}-2 \log ^2\frac{x}{z} \log \frac{y}{z}+\frac{2}{3} \log ^3\frac{x}{z}\right.\right.\\
&\left.\left.+\log ^2\frac{x}{z}+15 \log \frac{x}{z}+\frac{8}{3} \log ^3\frac{y}{z}-6 \log ^2\frac{y}{z}-28 \log \frac{y}{z}-\frac{10 \pi ^2}{3}+10\right)\right)\\
\end{split}
\label{eq.5.8}
\end{equation}
\begin{equation}
\begin{split}
&C_2^1=1+\frac{\alpha _s C_F}{4\pi}\left(6 \text{Li}_2(\frac{y}{x})+2 \log \frac{y}{z} \log \frac{x^2}{y z}+5 \log \frac{x z}{y^2}+\frac{y \log \frac{x}{y}}{x-y}+3 \log \frac{x y}{(x-y)^2} \log \frac{x}{y}\right.\\
&\left.-2 \pi ^2+\log\frac{z}{\mu^2}-1\right)\\
\end{split}
\label{eq.5.9}
\end{equation}
\begin{equation}
\begin{split}
&C_3=0\\
\end{split}
\label{eq.5.10}
\end{equation}
\begin{equation}
\begin{split}
&C_4^1=1+\frac{\alpha _s C_F}{4\pi}\left(-15+\frac{4y}{x-y}+\frac{6x\log\frac{x}{\mu^2}-2x}{y}+3 \log\frac{y}{\mu^2}\right.\\
&\left.-\left(2-\frac{2 x \left(4 x^2-7 x y+2 y^2\right)}{y (x-y)^2}\right) \text{Li}_2(1-\frac{x}{y})-(4x-2y)d_1\right.\\
&\left.+\text{Li}_2(1-\frac{y}{x}) \left(\frac{x \left(12 x^3-40 x^2 y+44 x y^2-17 y^3\right)}{y^2 (x-y)^2}-\frac{(4 x (2 x-y)) \log \frac{x}{y}}{y^2}\right)\right.\\
&\left.+\frac{(4 x (2 x-y)) \left(-\text{Li}_3(1-\frac{y}{x})+\text{Li}_3(\frac{y}{x})+\zeta (3)\right)}{y^2}\right.\\
&\left.+\left(\frac{x \left(4 x^2-27 x y+18 y^2\right)}{y (x-y)^2}+\frac{2 \pi ^2 x (2 x-y)}{3 y^2}-4\right) \log \frac{x}{y}\right.\\
&\left.+\frac{2\pi ^2 x(4 y-3 x)}{3y^2}+\left(\frac{2 x (2 x-y) \log \frac{x-y}{x}}{y^2}-\frac{3 x y}{2 (x-y)^2}\right) \log ^2\frac{x}{y}\right)\\
\end{split}
\label{eq.5.11}
\end{equation}
\begin{equation}
\begin{split}
&C_5^1=\frac{\alpha _s C_F}{4\pi}\left(-3 \log\frac{y}{\mu^2}+6+\left(\frac{3 y^2}{2 (x-y)^2}-\frac{2 x \log \frac{x-y}{x}}{y}\right) \log ^2\frac{x}{y}-\frac{x+3y}{x-y}\right.\\
&\left.+2yd_1+\frac{\left(6 x y-4 x^2\right) \text{Li}_2(1-\frac{x}{y})}{(x-y)^2}+\text{Li}_2(1-\frac{y}{x}) \left(\frac{-6 x^3+14 x^2 y-10 x y^2+3 y^3}{y (x-y)^2}+\frac{4 x\log \frac{x}{y}}{y}\right)\right.\\
&\left.-\frac{4 x \left(-\text{Li}_3(1-\frac{y}{x})+\text{Li}_3(\frac{y}{x})+\zeta (3)\right)}{y}+\frac{\pi ^2 \left(3 x^2-4 x y+y^2\right)}{3 y (x-y)}\right.\\
&\left.+\left(\frac{-5 x^2+8 x y+2 y^2}{(x-y)^2}-\frac{2 \pi ^2 x}{3 y}\right) \log \frac{x}{y}\right)\\
\end{split}
\label{eq.5.12}
\end{equation}
\begin{equation}
\begin{split}
&C_6=1\\
\end{split}
\label{eq.5.13}
\end{equation}
where $d_1$ is defined in Eq.~(\ref{eq.4.28}). Except for $C_6$, the other $C_i^j$'s denote the coefficients for $O_i$'s, and the products are at $O(\Lambda _{\rm QCD}\left./m_Q\right)^j$ order. It is now clear that the coefficient with large logarithm at order $O(1)$ is $C_1^0$. And the counter term of $C_1^0$ at the leading order is
\begin{equation}
\begin{split}
&Z_{C_1^0}=1+\frac{\alpha _s(\mu)C_F}{4\pi}\frac{2}{\epsilon}\\
\end{split}
\label{eq.5.14}
\end{equation}
which leads to the group function~(the anomalous dimension of the operator $O_1$ at the order of $O(\Lambda _{\rm QCD}/m_Q)^0$)
\begin{equation}
\begin{split}
&\gamma _{C_1^0}=-\frac{\alpha _s(\mu)C_F}{2\pi}\\
\end{split}
\label{eq.5.15}
\end{equation}

The effective field theory~(EFT) is often used to resume the large logarithms~\cite{SCET,Genon work}. In general, the coefficient can be written as
\begin{equation}
\begin{split}
&C(\mu)=H(\mu)\times J(\mu)\\
\end{split}
\label{eq.5.16}
\end{equation}
with $H$ known as the hard-function, and $J$ known as the jet function. Inspired by the idea of resummation using an EFT, we find we can separate the coefficient as a product of coefficients at different scale, as
\begin{equation}
\begin{split}
&C_1^0(\mu)=H(\mu)\times J_1(\mu)\times J_2(\mu)\\
\end{split}
\label{eq.5.18}
\end{equation}
with $H(\mu)$ is the hard-function and $J(\mu)=J_1(\mu)\times J_2(\mu)$ is the jet function. We find the RGE of $C(\mu)$
\begin{equation}
\begin{split}
&\mu \frac{\partial }{\partial \mu}C(\mu)=\gamma_C C(\mu)\\
\end{split}
\label{eq.5.19}
\end{equation}
becomes
\begin{equation}
\begin{split}
&\mu \frac{\partial }{\partial \mu} H(\mu)=\gamma_{H} H(\mu)\\
&\mu \frac{\partial }{\partial \mu} J_1(\mu)=\gamma_{J_1} J_1(\mu)\\
&\mu \frac{\partial }{\partial \mu} J_2(\mu)=\gamma_{J_2} J_2(\mu)\\
\end{split}
\label{eq.5.20}
\end{equation}
with
\begin{equation}
\begin{split}
&\gamma_C=\gamma_H+\gamma_{J_1}+\gamma_{J_2}
\end{split}
\label{eq.5.21}
\end{equation}
and
\begin{equation}
\begin{split}
&\gamma _H=-\frac{\alpha _s(\mu)C_F}{2\pi}\left(-1-\log\frac{z}{x}\right),\;\;\;\gamma _{J_1}=-\frac{\alpha _s(\mu)C_F}{2\pi}\left(1-\log \frac{x}{z}\right)\\
&\gamma _{J_2}=-\frac{\alpha _s(\mu)C_F}{2\pi}\\
\end{split}
\label{eq.5.23}
\end{equation}
We do not need the explicit form of $H(\mu)$ and $J_1(\mu)$. We can assume $H(\mu)$ is at scale $m_Q$, and $J_1(\mu)$ is at scale $\left.x/y\right.\sqrt{z}$, with $\left.x/y\right.\sqrt{z}\geq \sqrt{z}\sim \sqrt{m_Q\Lambda _{\rm QCD}}$, and let
\begin{equation}
\begin{split}
&H(m_Q)=1,\;\;\;J_1(\frac{x}{y}\sqrt{z})=1\\
\end{split}
\label{eq.5.22}
\end{equation}
With Eqs.~(\ref{eq.5.20}), (\ref{eq.5.23}) and (\ref{eq.5.22}), we solve the RGE evaluation of $H$ and $J_1$, the results are
\begin{equation}
\begin{split}
&\frac{H(\mu)}{H(m_Q)}=\left(\frac{\alpha _s(\mu)}{\alpha _s(m_Q)}\right)^{\frac{C_F}{\beta _0}\left(-1-\log \frac{z}{x}\right)}=1-\frac{\alpha _s(m_Q)C _f}{4\pi}\left(-\log\frac{\mu^2}{x}-\log \frac{\mu^2}{x}\log \frac{z}{x}\right)+O(\alpha _s^2)
\end{split}
\label{eq.5.24}
\end{equation}
\begin{equation}
\begin{split}
&\frac{J_1(\mu)}{J_1(\frac{x}{y}\sqrt{z})}=\left(\frac{\alpha _s(\mu)}{\alpha _s(\frac{x}{y}\sqrt{z})}\right)^{\frac{C_F}{\beta _0}\left(1-\log \frac{x}{z}\right)}=1-\frac{\alpha _s(\frac{x}{y}\sqrt{z})C_F}{4\pi}\left(\log\frac{\mu^2y^2}{x^2z}-\log \frac{\mu^2y^2}{x^2z}\log \frac{x}{z}\right)+O(\alpha _s^2)
\end{split}
\label{eq.5.27}
\end{equation}
where $\beta_0=11\left./3\right. C_A-2\left./3\right. n_f$.

Compare them with $C_1^0$, we find $J_2$ can be defined as
\begin{equation}
\begin{split}
&J_2(\mu)=1-\frac{\alpha _s(\mu) C_F}{4\pi}\left(6+\frac{\pi^2}{3}+\frac{y}{x}\log\frac{x}{y}-\log\frac{z}{\mu^2}-2{\rm Li}_2\frac{y}{x}+2\log\frac{x}{y}\log\frac{x-y}{x}\right)\\
\end{split}
\label{eq.5.28}
\end{equation}
Then the resummed $C_1^0$ can be written as
\begin{equation}
\begin{split}
&C_{1r}^0(\mu)=\left(\frac{\alpha _s(\mu)}{\alpha _s(m_Q)}\right)^{\frac{C_F}{\beta _0}\left(-1-\log \frac{z}{x}\right)}\times \left(\frac{\alpha _s(\mu)}{\alpha _s(\frac{x}{y}\sqrt{z})}\right)^{\frac{C_F}{\beta _0}\left(1-\log \frac{x}{z}\right)}\times J_2(\mu)\\
\end{split}
\label{eq.5.29}
\end{equation}
where $C_{1r}^0$ denotes the resummed $C_1^0$. There is no longer large logarithms in $J_2(\mu)$ when $\mu=\sqrt{z}\sim \sqrt{m_Q\Lambda _{\rm QCD}}$. And
\begin{equation}
\begin{split}
&C_{1r}^0(\sqrt{z})=C_1^0(\sqrt{z})+O(\alpha _s^2)
\end{split}
\label{eq.5.30}
\end{equation}
which means that, if the resummed result of the coefficient $C_{1r}^0$ is expanded to the leading order at scale $\mu=\sqrt{z}$, it will go back to the coefficient $C_1^0$ as expected.

We also notice that, when $E_{\gamma}$ becomes smaller, for example, when $E_{\gamma} \sim \sqrt{m_Q\Lambda _{\rm QCD}}$, $C_{1r}^0(\mu)$ can still correctly resume all the large logarithms in the leading order in $C_1^0(\mu)$ because all the remaining $\log \left( x / y\right)$ terms will go to sub-leading order.

\subsection{\label{sec:52}the soft photon region}

In soft photon region that $E_{\gamma} \sim \Lambda _{\rm QCD}^2 \left./m_Q\right.$, we find
\begin{equation}
\begin{split}
&C_1^1=1-\frac{\alpha _s C_F}{4\pi}\left(12-\frac{8\pi^2}{3}-\log\frac{z}{\mu^2}-5\log\frac{w^2}{xz} +\log^2\frac{w^2}{xz}\right.\\
&\left.+\frac{2xz}{wy}\left(-5 + 3\pi^2 +5 \log\frac{w^2}{xz}+2\log\frac{z}{w}\log\frac{w^2}{xz}\right)\right.\\
&\left.+ 2\pi i \left(2\log\frac{xz}{w^2} +5+\frac{2 x z}{w y}(-2\log\frac{xz}{w^2} -5+\log\frac{x}{z})\right)\right)\\
\end{split}
\label{eq.5.31}
\end{equation}
\begin{equation}
\begin{split}
&C_2^0=1+\frac{\alpha _s C_F}{4\pi}\left(-7+\log\frac{z}{\mu^2} -\frac{\pi^2}{3} + \log\frac{z}{x}\log\frac{xz}{w^2} - \log\frac{xz}{w^2}+ 2\pi i\left(-1-\log\frac{x}{z}\right)\right)\\
\end{split}
\label{eq.5.32}
\end{equation}
\begin{equation}
\begin{split}
&C_2^1=\frac{\alpha _s C_F}{4\pi}\left(\frac{4w\log \frac{w}{x}}{x}+\frac{y}{w}\left(-2 \log \frac{w}{z} \log \frac{w^2}{x z}+8 \log \frac{w^2}{x z}+4 \log \frac{z}{w}+3 \pi ^2-8\right)\right.\\
&\left.+\frac{x z}{w^2}\left(\left(\frac{2}{3} \log \frac{w}{x} \log \frac{x}{z}+2 \log \frac{w}{x}+\frac{4}{3} \log ^2\frac{w}{z}-3 \log \frac{x}{z}-12\right) \log \frac{w^2}{x z}\right.\right.\\
&\left.\left.+\frac{\pi ^2}{3}\left(2 \log \frac{x}{w}+18 \log \frac{z}{w}\right)+4 \zeta (3)-\frac{4 \pi ^2}{3}+2\right)\right.\\
&\left.+2 \pi i \left(-\frac{y \left(\log \frac{x z^3}{w^4}+6\right)}{w}-\frac{x z \left(4 \log \frac{w}{x}+4 \log ^2\frac{w}{z}-\log ^2\frac{x}{z}+2 \log \frac{z}{x}-\frac{2 \pi ^2}{3}-12\right)}{w^2}-\frac{2 w}{x}\right)\right)\\
\end{split}
\label{eq.5.33}
\end{equation}
\begin{equation}
\begin{split}
&C_3^0=\frac{\alpha _s C_F}{4\pi}\left(-\frac{2x}{y}\left(4-3\log\frac{x}{\mu^2}\right)\right)
\end{split}
\label{eq.5.34}
\end{equation}
\begin{equation}
\begin{split}
&C_3^1=1+\frac{\alpha _s C_F}{4\pi}\left(-7+3\log\frac{x}{\mu^2}-10\log\frac{x}{y}\right)
\end{split}
\label{eq.5.35}
\end{equation}
\begin{equation}
\begin{split}
&C_4^1=\frac{\alpha _s C_F}{4\pi}\left(-\frac{2x}{y}\left(4-3\log\frac{x}{\mu^2}\right)\right)\\
\end{split}
\label{eq.5.36}
\end{equation}
\begin{equation}
\begin{split}
&C_5^0=\frac{\alpha _s C_F}{4\pi}\left(4-3\log\frac{x}{\mu^2}\right)\\
\end{split}
\label{eq.5.37}
\end{equation}
\begin{equation}
\begin{split}
&C_6=1\\
\end{split}
\label{eq.5.38}
\end{equation}

Then we concentrate on the resummation of the large logarithms in $C_2^0$. The anomalous dimension for $O_2$ at leading order is found to be
\begin{equation}
\begin{split}
&\gamma _{C_2^0}=-\frac{\alpha _s(\mu)C_F}{2\pi}\\
\end{split}
\label{eq.5.39}
\end{equation}
Similar as the hard photon case, we assume
\begin{equation}
\begin{split}
&C_2^0(\mu)=h(\mu)\times j_1(\mu)\times j_2(\mu)\\
\end{split}
\label{eq.5.40}
\end{equation}
and the scales are
\begin{equation}
\begin{split}
&h(\sqrt{\frac{w^3}{xz}})=1,\;\;\;j_1(\sqrt{w})=1\\
&\sqrt{\frac{w^3}{xz}}\sim m_Q,\;\;\;\sqrt{w} \sim \sqrt{m_Q\Lambda _{\rm QCD}}
\end{split}
\label{eq.5.41}
\end{equation}
Similar to Eq.~(\ref{eq.5.21}), we separate the anomalous dimension as
\begin{equation}
\begin{split}
&\gamma _h=\frac{\alpha _s(\mu)C_F}{2\pi}\left(1-\log\frac{x}{z}+6\pi i)\right),\;\;\;\gamma _{j_1}=\frac{\alpha _s(\mu)C_F}{2\pi}\left(-1+\log\frac{x}{z}-6\pi i\right)\\
&\gamma _{j_2}=-\frac{\alpha _s(\mu)C_F}{2\pi}\\
\end{split}
\label{eq.5.42}
\end{equation}

The RGE evaluation of $h(\mu)$ is found to be
\begin{equation}
\begin{split}
&\frac{h(\mu)}{h(\sqrt{\frac{w^3}{xz}})}=\left(\frac{\alpha _s(\mu)}{\alpha _s(\sqrt{\frac{w^3}{xz}})}\right)^{-\frac{C_F}{\beta _0}\left(1-\log \frac{x}{z}+6\pi i\right)}\\
&=1+\frac{\alpha _s(\sqrt{\frac{w^3}{xz}})C _F}{4\pi}\left((1+6\pi i)\log\frac{\mu^2xz}{w^3}+\log \frac{\mu^2xz}{w^3}\log \frac{z}{x}\right)+O(\alpha _s^2)
\end{split}
\label{eq.5.43}
\end{equation}
and the result of $j_1(\mu)$ is
\begin{equation}
\begin{split}
&\frac{j_1(\mu)}{j_1(\sqrt{w})}=\left(\frac{\alpha _s(\mu)}{\alpha _s(\sqrt{w})}\right)^{-\frac{C_F}{\beta _0}\left(-1-\log \frac{z}{x}-6\pi i\right)}\\
&=1+\frac{\alpha _s(\sqrt{w})C_F}{4\pi}\left((-1-6\pi i)\log\frac{\mu^2}{w}-\log \frac{\mu^2}{w}\log \frac{z}{x}\right)+O(\alpha _s^2)
\end{split}
\label{eq.5.44}
\end{equation}
and $j_2$ can be obtained as
\begin{equation}
\begin{split}
&j_2=1+\frac{\alpha _s(\mu)C_F}{4\pi}\left(-7-\frac{\pi^2}{3}+\log\frac{w^4y}{x^3z^2}-\log\frac{\mu^2y}{xz}+ 2\pi i(-1+2\log\frac{w^3}{x^2z})\right)\\
\end{split}
\label{eq.5.45}
\end{equation}
where one can see that the scale of $j_2$ is at $\sqrt{\left.xz/y\right.}\sim \sqrt{m_Q\Lambda _{\rm QCD}}$. At this scale, all the logarithm terms are small.

The resummed $C_{2r}^0(\mu)$ is
\begin{equation}
\begin{split}
&C_{2r}^0(\mu)=\left(\frac{\alpha _s(\mu)}{\alpha _s(\sqrt{\frac{wy}{z}})}\right)^{-\frac{C_F}{\beta _0}\left(1-\log \frac{x}{z}+6\pi i\right)} \times \left(\frac{\alpha _s(\mu)}{\alpha _s(\sqrt{w})}\right)^{-\frac{C_F}{\beta _0}\left(-1-\log \frac{z}{x}-6\pi i\right)}\times j_2(\mu)
\end{split}
\label{eq.5.46}
\end{equation}
and then evaluate $h(\mu)$ and $j_1(\mu)$ to $\sqrt{\left.xz/y\right.}$, considering $\sqrt{w} \geq \sqrt{\left.xz/y\right.}\sim \sqrt{m_Q\Lambda _{\rm QCD}}$, we find
\begin{equation}
\begin{split}
&C_{2r}^0(\sqrt{\frac{xz}{y}})=C_2^0(\sqrt{\frac{xz}{y}})+O(\alpha _s^2)
\end{split}
\label{eq.5.47}
\end{equation}
The remaining $\gamma _{j_2}$ can reproduce the $\log \left( \mu^2 y/xz\right)$ term in $j_2(\mu)$. Notice that, the scale $\sqrt{\left.xz/y\right.}$ does not depend on $E_{\gamma}$. And the factorization scale for the case of the photon being soft is at $\sqrt{m_Q\Lambda _{\rm QCD}}$.

\subsection{\label{sec:53}the other large terms in the soft photon region}

Apart from the large logarithms in the weak vertex correction, there are also large terms in $C_3$, $C_4$ and $C_5$, all with the form
\begin{equation}
\begin{split}
&3\log\frac{x}{\mu^2}-4
\end{split}
\label{eq.5.48}
\end{equation}
They all come from $F_b^{(1)\rm Wfc}$, which implies that the relevant scale is at
\begin{equation}
\begin{split}
&\mu=\frac{m_Q}{e^{\frac{2}{3}}}\approx \frac{m_Q}{2}
\end{split}
\label{eq.5.49}
\end{equation}
We will use this scale in the numerical calculation.

\section{\label{sec:numerical}Numerical Results}

\begin{figure}
\includegraphics[scale=0.65]{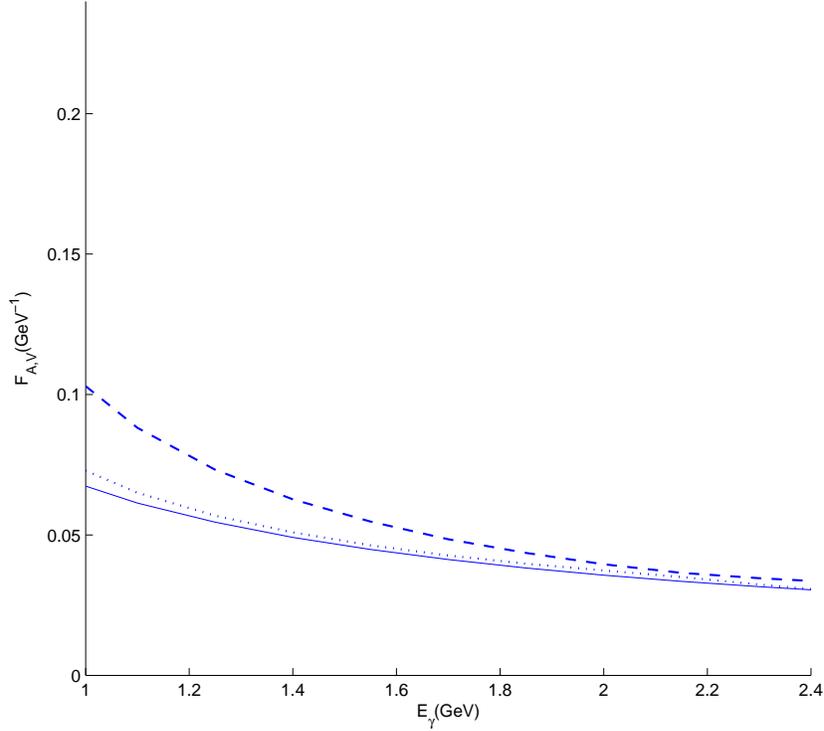}
\caption{\label{b_form_largeE} Form factors of $B\to \gamma e \nu _e$. The solid line is for the 1-loop result at the leading order of $\Lambda _{\rm QCD}\left/m_Q\right.$. The dotted line is for the result of $F_V$, while the dashed line is the result of $F_A$.}
\end{figure}

The contribution of Fig.~\ref{Fig1:feynman_tree}.c depends on not only $E_{\gamma}$ but also on $p_{\nu}$ and $p_l$. For simplicity, we treat this term separately, then the form factors are defined as
\begin{equation}
\begin{split}
&\langle \gamma\mid \bar{q} \Gamma ^{\mu} Q\mid P\rangle+F_c p_P^{\mu}=\epsilon _{\mu\nu\rho\sigma}\varepsilon ^{\nu}p_P^{\rho}p_{\gamma}^{\sigma}(F_V+F_c)\\
&+i\left(\varepsilon ^{\mu}p_P\cdot p_{\gamma}-p_{\gamma}^{\mu}\varepsilon \cdot p_P\right)(F_A+F_c)+2(p_l\cdot \varepsilon _{\gamma})F_c p_P^{\mu}\\
\end{split}
\label{eq.6.1}
\end{equation}
When $E_{\gamma} \sim m_Q$, up to sub-leading order, we find
\begin{equation}
\begin{split}
&F_V=\frac{1}{(2\pi)^3}\frac{3}{\sqrt{6}}\int d^3 k \Psi (k)\frac{1}{2\sqrt{p_{q0}p_{Q0}(p_{q0}+m_q)(p_{Q0}+m_Q)}} \frac{1}{m_PE_{\gamma}}\\
&\times \left(e_q(2m_Q-p_{q0})C_1+2e_Q(p_{q0}-p_{q3})C_4+e_Q\frac{2m_Qp_{q3}}{E_{\gamma}}C_5\right)\\
&F_A=\frac{1}{(2\pi)^3}\frac{3}{\sqrt{6}}\int d^3 k \Psi (k)\frac{1}{2\sqrt{p_{q0}p_{Q0}(p_{q0}+m_q)(p_{Q0}+m_Q)}} \frac{1}{m_PE_{\gamma}}\\
&\times \left(e_q(2m_Q-p_{q0})C_1-e_q\frac{2m_Q(p_{q0}-p_{q3})}{E_{\gamma}}C_2-2e_Q(p_{q0}-p_{q3})C_4-e_Q\frac{2m_Qp_{q0}}{E_{\gamma}}C_5\right)\\
\end{split}
\label{eq.6.2}
\end{equation}
When $E_{\gamma} \sim \left. \Lambda^2 _{\rm QCD}/ m_Q\right.$, up to sub-leading order, we find
\begin{equation}
\begin{split}
&F_V=\frac{1}{(2\pi)^3}\frac{3}{\sqrt{6}}\int d^3 k \Psi (k)\frac{1}{2\sqrt{p_{q0}p_{Q0}(p_{q0}+m_q)(p_{Q0}+m_Q)}} \frac{1}{m_PE_{\gamma}}\\
&\times \left(2e_qm_QC_1\right)\\
&F_A=\frac{1}{(2\pi)^3}\frac{3}{\sqrt{6}}\int d^3 k \Psi (k)\frac{1}{2\sqrt{p_{q0}p_{Q0}(p_{q0}+m_q)(p_{Q0}+m_Q)}} \frac{1}{m_PE_{\gamma}}\\
&\times \left(2e_qm_QC_1-e_q\frac{(2m_Q-p_{q0})(p_{q0}-p_{q3})}{E_{\gamma}}C_2-e_Q\frac{2(p_{q0}^2-p_{q3}^2)}{E_{\gamma}}C_3\right)\\
\end{split}
\label{eq.6.3}
\end{equation}
In the numerical calculation, the values of parameters we take are~\cite{Teacher Yang wave function}
\begin{equation}
\begin{split}
&m_D=1.9\;{\rm GeV},\;m_B=5.1\;{\rm GeV},\;m_u=m_d=0.08\;{\rm GeV},\;m_b=4.98\;{\rm GeV},\;m_c=1.54\;{\rm GeV}\\
&\Lambda _{\rm QCD}=200\;{\rm MeV},\;\lambda _B = 2.8\;{\rm GeV}^{-1},\;\lambda _D = 3.4\;{\rm GeV}^{-1}%,\lambda _{D_s} = 3.2{\rm GeV}^{-1}
\end{split}
\label{eq.6.4}
\end{equation}
For $B$ meson, and for the hard photon region, we calculate the numerical results of form factors with $E_{\gamma}>m_Q\left.  / \right. 4$. The result is shown in Fig.~\ref{b_form_largeE}.
Comparing with Fig.~7 in Ref.~\cite{Our work}, we find that, at the large $E_{\gamma}$ region, the numerical result of the form factor dose not change very much. Both $F_A$ and $F_V$ only increase slightly.

\begin{figure}
\includegraphics[scale=0.65]{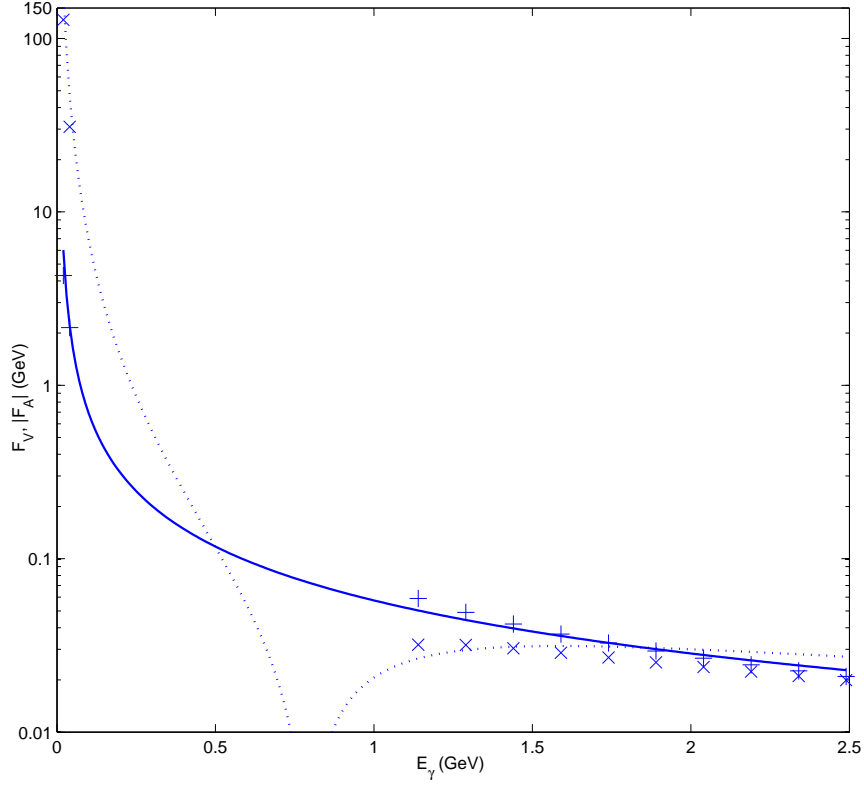}
\caption{\label{fit_b_tree} Fit of the form factors of $B\to \gamma e \nu _e$ at tree level. The `$\times$' points and the dotted line are for $|F_A|$, while the solid line and the `+' points are for $F_V$. $F_A$ will go from positive to negative, and therefor not possible to show in a $\log$ scale figure, so we show $|F_A|$ instead.}
\end{figure}

In the soft photon region, the form factors at leading order is very different from the hard photon region even at tree level. In the soft photon region, $F_A$ will receive $O\left(1/E_{\gamma}\right)$ contribution, while $F_V$ is at the sub-leading order $O(1/\left.\Lambda _{\rm QCD}\right)$. The relation $F_A=F_V$ \cite{FaneqFv} is broken at leading order. So we investigate the tree level result at first. The numerical results of the form factors are inconvenient to use when calculate the decay widths. For simplicity, we use some simple forms to fit the numerical results, inspired by the form factors given in Ref.~\cite{naive factorization}, the form factors are fitted as
\begin{equation}
\begin{split}
&F=\left(A\frac{\Lambda_{\rm QCD}}{E_{\gamma}}+B\left(\frac{\Lambda_{\rm QCD}}{{E_{\gamma}}}\right)^2\right)
\end{split}
\label{eq.6.5}
\end{equation}
The fitted result of the tree level form factors is shown in Fig.~\ref{fit_b_tree}, which can be written as
\begin{equation}
\begin{split}
&F^{B\to \gamma l\nu}_{A\;tree}(E_{\gamma})=\left(0.50\frac{\Lambda_{\rm QCD}}{E_{\gamma}}-1.97\left(\frac{\Lambda_{\rm QCD}}{{E_{\gamma}}}\right)^2\right)\\
&F^{B\to \gamma l\nu}_{V\;tree}(E_{\gamma})=\left(0.28\frac{\Lambda_{\rm QCD}}{E_{\gamma}}+0.032\left(\frac{\Lambda_{\rm QCD}}{{E_{\gamma}}}\right)^2\right)\\
\end{split}
\label{eq.6.6}
\end{equation}

In the calculation, we treat the 3-momentum of the light-quark $k$ as small quantity, i.e. $k\sim \Lambda _{\rm QCD}$. For convenience, we use the average value of the light-quark momentum $\langle k \rangle$ in the meson as a small scale instead of $\Lambda _{\rm QCD}$. In the soft photon region, we calculate the form factors at the region $\left.\langle k\rangle ^2/ m_Q\right. > E_{\gamma}>\left.2 \langle k\rangle^2 / m_Q\right.$. Using the wave function given in Eq.~(\ref{eq.5.2}), we find, for $B$ meson, $\langle k\rangle=0.357\;{\rm GeV}$.

The result that $F_A$ will go from positive to negative when $E_{\gamma}$ become small can also been found in Eq.~(2.9) of Ref.~\cite{Beneke work}, because the signs of the $1/E_{\gamma}^2$ term and $1/E_{\gamma}$ terms are different in $F_A$.

\begin{figure}
\includegraphics[scale=0.65]{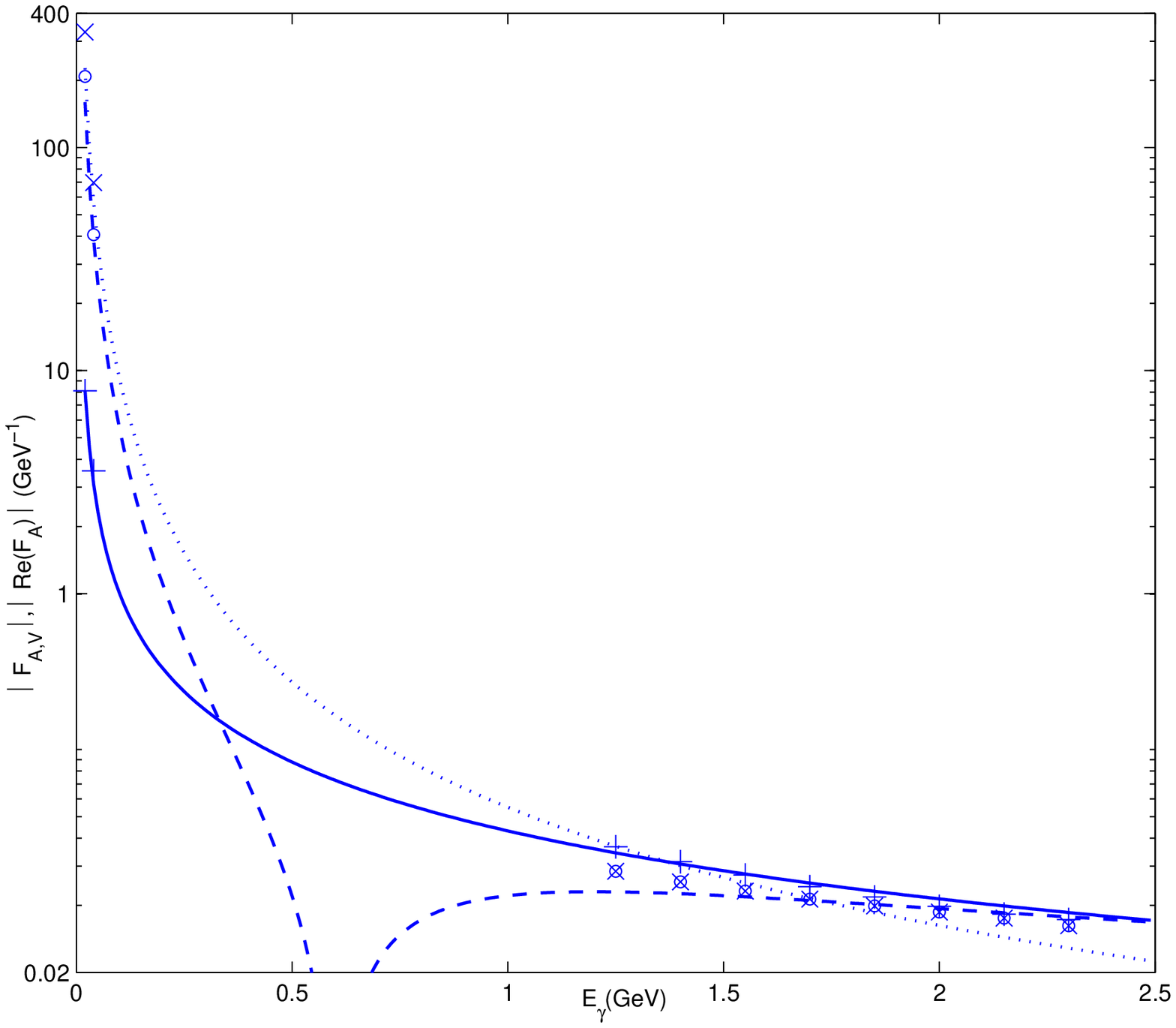}
\caption{\label{fit_b} Fit of the form factors of $B\to \gamma e \nu _e$ at 1-loop level. The `$\times$' points and the dotted line are for $|F_A|$, while the solid line and the `+' points are for $|F_A|$. $\mathcal{R}e(F_A)$ will go from positive to negative, and therefor not possible to show in a $\log$ scale figure, so we show $|\mathcal{R}e (F_A)|$ instead. The dashed line and the 'o' points are for $|\mathcal{R}e (F_A)|$.}
\end{figure}

To calculate the decay width, we need the result of $|F_A|$, $|F_B|$ and $\mathcal{R}e (F_A)$. The fit of the form factors of $B$ meson at one-loop level is shown in Fig.~\ref{fit_b}, and the fitted results can be written as
\begin{equation}
\begin{split}
&|F^{B\to \gamma l\nu}_A(E_{\gamma})|=\left(0.14\frac{\Lambda_{\rm QCD}}{E_{\gamma}}+2.21\left(\frac{\Lambda_{\rm QCD}}{{E_{\gamma}}}\right)^2\right)\\
&|F^{B\to \gamma l\nu}_V(E_{\gamma})|=\left(0.42\frac{\Lambda_{\rm QCD}}{E_{\gamma}}+0.032\left(\frac{\Lambda_{\rm QCD}}{{E_{\gamma}}}\right)^2\right)\\
&\mathcal{R}e (F^{B\to \gamma l\nu}_A(E_{\gamma}))=\left(0.56\frac{\Lambda_{\rm QCD}}{E_{\gamma}}-1.71\left(\frac{\Lambda_{\rm QCD}}{{E_{\gamma}}}\right)^2\right)\\
\end{split}
\label{eq.6.7}
\end{equation}
On the other hand, $F_c$ can be related to the decay constant \cite{Teacher Yang wave function,naive factorization} by
\begin{equation}
\begin{split}
&F_c p_P^{\mu}=\frac{1}{p_l\cdot p_{\gamma}}\langle 0|\bar{q}\gamma _{\mu}(1-\gamma_5)Q|P\rangle= \frac{-f_Pp_P^{\mu}}{p_l\cdot p_{\gamma}}
\end{split}
\label{eq.6.8}
\end{equation}
Using the fitted results of $F_A$, $F_V$ and $F_c$, and using the Cabibbo - Kobayashi - Maskawa (CKM) matrix elements~\cite{particaldatagroup} and the decay constant $f_B$~\cite{Teacher Yang wave function} as following
\begin{equation}
\begin{split}
&f_B=0.194\;{\rm GeV^{-1}},\;V_{\rm ub}=0.00413\\
\end{split}
\label{eq.6.9}
\end{equation}
we can obtain the branching ratios.

There are IR divergences in the radiative leptonic decays in the case that the photon is soft or the photon is collinear with the emitted lepton. Theoretically this IR divergences can be canceled by adding the decay rate of the radiative leptonic decay with the pure leptonic decay rate with one-loop virtual photon correction~\cite{changch}. The radiative leptonic decay can not be distinguished from the pure leptonic decay in experiment when the photon energy is smaller than the experimental resolution to the photon energy. So the decay rate of the radiative leptonic decay depend on the experimental resolution to the photon energy which is denoted by $\Delta E_{\gamma}$.

Using $\Delta E_{\gamma} = 10{\rm MeV}$ \cite{photon res}, we obtain the branch ratios of $B\to \gamma e\nu_{e}$. We also calculate the tree level result neglecting the contribution of orders $\Lambda _{\rm QCD}\left./\right.E_{\gamma}$ and $\Lambda _{\rm QCD}\left./\right.m_Q$ for comparison, both the tree level and up-to one-loop level results of the branching ratios are given below
\begin{equation}
\begin{split}
&{\rm Br}_{\left(\frac{\alpha_s\Lambda _{\rm QCD}}{E_{\gamma}}\right)^0}(B\to \gamma e\nu_{e})=1.73\times 10^{-6}\\
&{\rm Br}_{\alpha_s^0}(B\to \gamma e\nu_{e})=2.51\times 10^{-6}\\
&{\rm Br}_{\left(\frac{\alpha_s\Lambda _{\rm QCD}}{E_{\gamma}}\right)^1}(B\to \gamma e\nu_{e})=5.21\times 10^{-6}\\
\end{split}
\label{eq.6.10}
\end{equation}
where ${\rm Br}_{\left(\frac{\alpha_s\Lambda _{\rm QCD}}{E_{\gamma}}\right)^0}$ is the tree level result calculated by neglecting the contribution of orders $\Lambda _{\rm QCD}\left./\right.E_{\gamma}$ and $\Lambda _{\rm QCD}\left./\right.m_Q$, ${\rm Br}_{\alpha_s^0}(B\to \gamma e\nu_{e})$ is the tree level result including those contribution and ${\rm Br}_{\left(\frac{\alpha_s\Lambda _{\rm QCD}}{E_{\gamma}}\right)^1}$ is the one-loop result.
From Eq.~(\ref{eq.6.10}), one can find that, both at tree and one-loop level, the branching ratio increases largely after including the soft-photon contribution. This result implies the soft photon region could be very important for $B$ mesons even the mass of $b$ quark is very large.

\begin{figure}
\includegraphics[scale=0.65]{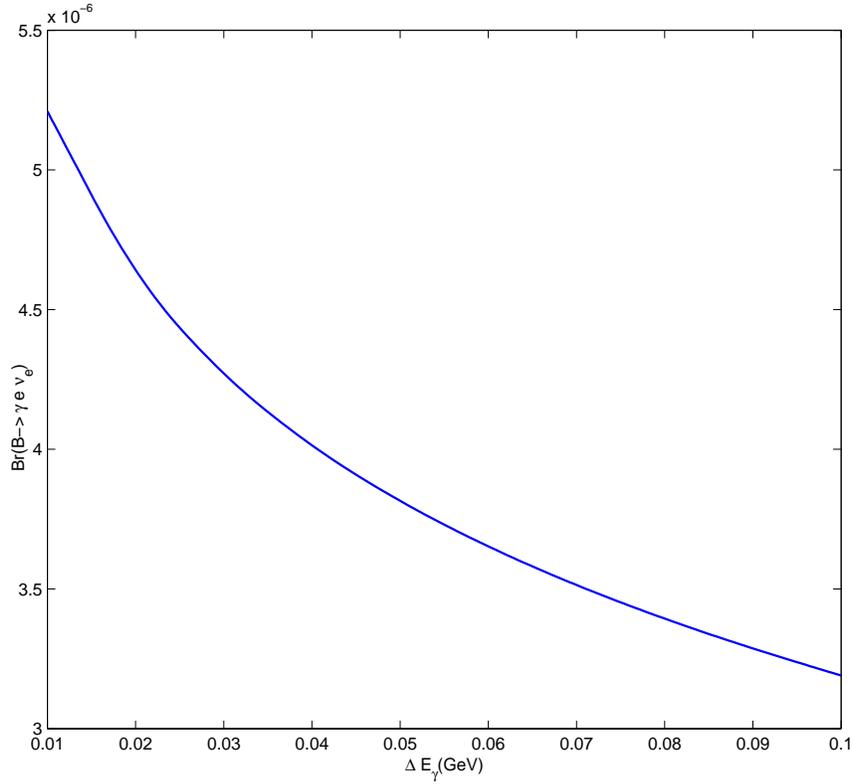}
\caption{\label{brBonDE} The branching ratio of $B\to \gamma e \nu_e$ at 1-loop level as a function of the resolution of photon energy $\Delta E_{\gamma}$.}
\end{figure}

The dependence of the branching ratios at 1-loop level on the resolution of photon energy is shown in Fig.~\ref{brBonDE} and Table~\ref{table:brBonDE}. The search for the radiative leptonic decay $B^+\to \ell^+\nu_{\ell}\gamma$ was performed by {\sl BABAR} Collaboration in 2009 \cite{BABAR2009} and Belle Collaboration in 2015 \cite{Belle2015}. The upper limit for the branching ratio of the radiative leptonic decay was presented. The upper limit given by {\sl BABAR} collaboration is $BR(B\to e\nu _e \gamma)<17\times 10^{-6}$ for the photon threshold energy $E_\gamma >20 {\rm MeV}$ \cite{BABAR2009}, while the result of Belle Collaboration is $BR(B\to e\nu _e \gamma)<6.1\times 10^{-6}$ for the photon threshold $E_\gamma >1 {\rm GeV}$, and $BR(B\to e\nu _e \gamma)<9.3\times 10^{-6}$ for the photon threshold $E_\gamma >400 {\rm MeV}$ \cite{Belle2015}. Compared with the experimental upper limit, our results given in Fig.~\ref{brBonDE} and Table~\ref{table:brBonDE} are consistent with the experimental data.
\begin{table}[ht]
\begin{tabular}{ c c| c c }
\hline
    $\Delta E_{\gamma}$ & $BR(B\to e\nu _e \gamma)$ & $\Delta E_{\gamma}$ & $BR(B\to e\nu _e \gamma)$\\
\hline
  $5{\rm MeV}$ & $5.81\times 10^{-6}$ & $20{\rm MeV}$ & $4.61\times 10^{-6}$ \\
  $10{\rm MeV}$ & $5.21\times 10^{-6}$ & $25{\rm MeV}$ & $4.42\times 10^{-6}$ \\
  $15{\rm MeV}$ & $4.86\times 10^{-6}$ & $30{\rm MeV}$ & $4.26\times 10^{-6}$ \\
  \hline
\end{tabular}
\caption{The branching ratios of $B\to \gamma e \nu_e$ at 1-loop level with different photon resolution $\Delta E_{\gamma}$.}
\label{table:brBonDE}
\end{table}

For $D$ mesons we find $\langle k\rangle=0.294\; {\rm GeV}$, and the energy of photon is $0\leq E_{\gamma} \leq \left.m_Q / 2\right. = 0.77{\rm GeV}$. $E_{\gamma\; \rm Max}$ is only less then $3$ times of $\langle k\rangle$. So, in the hard photon region, we calculate the numerical results with $E_{\gamma}>m_Q\left.  / \right. 2 - \Lambda _{\rm QCD}$, and in the soft photon region, we calculate in the region $\left.\langle k\rangle^2/ m_Q\right. >E_{\gamma}>\left.2 \langle k\rangle^2 / m_Q\right.$.

\begin{figure}
\includegraphics[scale=0.65]{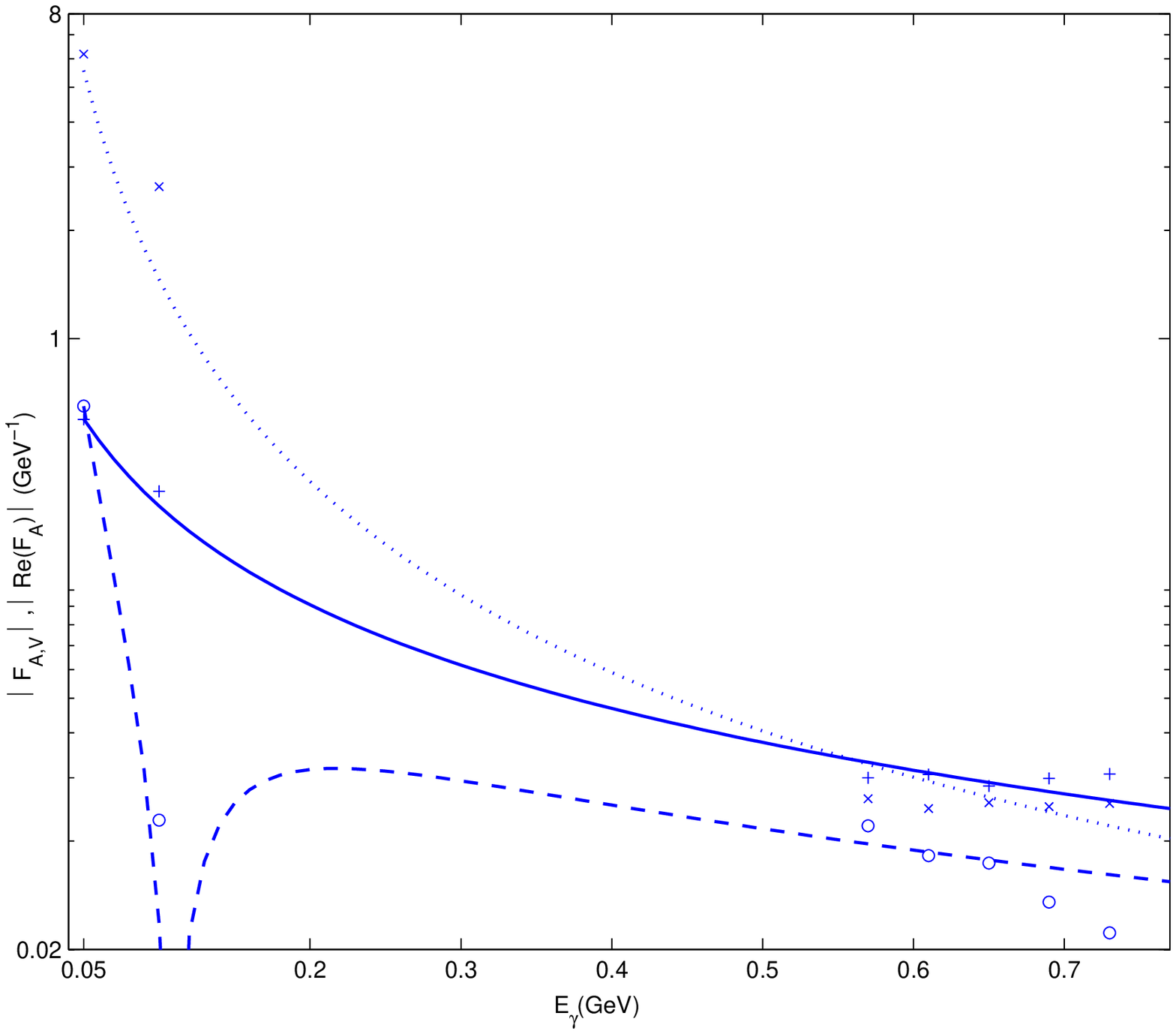}
\caption{\label{fit_d} Fit of the form factors of $D\to \gamma e \nu _e$ at 1-loop level. The `$\times$' points and the dotted line are for $|F_A|$ while the solid line and the `+' points are for $|F_A|$. $\mathcal{R}e(F_A)$ will go from positive to negative, and therefor not possible to show in a $\log$ scale figure, so we show $|\mathcal{R}e (F_A)|$ instead. The dashed line and the 'o' points are for $|\mathcal{R}e (F_A)|$.}
\end{figure}

The fit of the form factors of $D\to \gamma e\nu_e$ at 1-loop level are shown in Fig.~\ref{fit_d}, and the fitted results can be written as
\begin{equation}
\begin{split}
&|F^{D\to \gamma l\nu}_A(E_{\gamma})|=\left(0.071\frac{\Lambda_{\rm QCD}}{E_{\gamma}}+0.33\left(\frac{\Lambda_{\rm QCD}}{{E_{\gamma}}}\right)^2\right)\\
&|F^{D\to \gamma l\nu}_V(E_{\gamma})|=\left(0.19\frac{\Lambda_{\rm QCD}}{E_{\gamma}}-0.011\left(\frac{\Lambda_{\rm QCD}}{{E_{\gamma}}}\right)^2\right)\\
&\mathcal{R}e (F^{D\to \gamma l\nu}_A(E_{\gamma}))=\left(0.14\frac{\Lambda_{\rm QCD}}{E_{\gamma}}-0.075\left(\frac{\Lambda_{\rm QCD}}{{E_{\gamma}}}\right)^2\right)\\
\end{split}
\label{eq.6.11}
\end{equation}

Using \cite{Teacher Yang wave function,particaldatagroup}
\begin{equation}
\begin{split}
&f_D=0.205\;{\rm GeV^{-1}},\;V_{\rm cd}=0.225\\
\end{split}
\label{eq.6.12}
\end{equation}

We obtain the branching ratios of $D\to \gamma e\nu_e$, and the branching ratios given with the dependence on $\Delta E_{\gamma}$ are shown in Table.~\ref{table:brDonDE}.
\begin{table}[ht]
\begin{tabular}{ c c| c c }
\hline
    $\Delta E_{\gamma}$ & $BR(D\to e\nu _e \gamma)$ & $\Delta E_{\gamma}$ & $BR(D\to e\nu _e \gamma)$\\
\hline
  $5{\rm MeV}$ & $2.03\times 10^{-5}$ & $20{\rm MeV}$ & $1.80\times 10^{-5}$ \\
  $10{\rm MeV}$ & $1.92\times 10^{-5}$ & $25{\rm MeV}$ & $1.76\times 10^{-5}$ \\
  $15{\rm MeV}$ & $1.85\times 10^{-5}$ & $30{\rm MeV}$ & $1.73\times 10^{-5}$ \\
  \hline
\end{tabular}
\caption{The branching ratios of $D\to \gamma e \nu_e$ at 1-loop level with different photon resolution $\Delta E_{\gamma}$.}
\label{table:brDonDE}
\end{table}

We find that, for $D$ mesons, the soft photon region is also very important. Compared with the tree level result of Ref.~\cite{Our work} where the contribution of order $\Lambda _{\rm QCD}\left. / \right. E_{\gamma}$ is neglected, the branching ratio is increased about $2$ times. However, the enhancement is not so large compared to the $B$ meson case. We find that it is because of the contribution of $C_3O_3$ in Eq.~(\ref{eq.6.3}). $C_3O_3$ is a $\left. \Lambda _{\rm QCD}^2/ E_{\gamma} m_Q \right.$ order contribution, while the contribution of $C_2O_2$ is at the order of $\left. \Lambda _{\rm QCD}/ E_{\gamma}\right.$. For $B$ meson, the mass of $b$ quark is very large, so the contribution of $C_3O_3$ compared to $C_2O_2$ is small. However, for $D$ meson, with the charge of the quark, we find
\begin{equation}
\begin{split}
&\left|e_Q\frac{\langle k\rangle^2}{m_QE_{\gamma}}\right| : \left|e_q\frac{\langle k\rangle }{E_{\gamma}}\right| \sim 39\%
\end{split}
\label{eq.6.13}
\end{equation}
As a result, the $\left.\Lambda _{\rm QCD}/E_{\gamma}\right.$ contribution could be canceled a lot by $C_3O_3$ which is relatively a $\left.\Lambda _{\rm QCD}/m_Q\right.$ order contribution.

\section{\label{sec:summary}Summary}

In this paper, we study the factorization of the radiative leptonic decays of $B^-$ and $D^-$ mesons. Compared with the work in Ref.~\cite{Our work}, the factorization is extended to include the $O(\Lambda _{\rm QCD}\left/E_{\gamma}\right.)$ contributions. The factorization is proved at 1-loop order, and the IR divergences are found to be absorbed into the wave-function at any order of $\Lambda _{\rm QCD}\left./m_Q\right.$ explicitly. The hard kernel at order $O(\alpha _s)$ is obtained. We list the hard kernel at order $O(\alpha _s \Lambda _{\rm QCD}\left./m_Q\right.)$ explicitly in both the soft photon region and the hard photon region. We use the wave function obtained in Ref.~\cite{Teacher Yang wave function} to calculate the numerical results. The branching ratio of $B^-\to \gamma e\bar{\nu}$ is found to be at the order of $5.21\times 10^{-6}$, which is a little larger then the previous works \cite{naive factorization,TM Yan work,prework1,prework2,prework3}. The results for the $D$ mesons radiative leptonic decay in the previous works are in the range from $10^{-3}$ to $10^{-6}$, our results is about $1.92\times 10^{-5}$.

We also find that the contribution of the soft photon region is very important especially for $B$ meson, the branching ratio is increased to be about $3$ times of the result in Ref.~\cite{Our work}, which is obtained using factorization and including $\Lambda _{\rm QCD}\left./\right.m_Q$ contributions but treating $E_{\gamma}\sim m_Q$. This result implies the importance of $O(\Lambda _{\rm QCD}\left/E_{\gamma}\right.)$ contributions, which increases the branching ratios at the soft photon region no matter how large the heavy quark mass is. For $D\to \gamma e\nu _e$, we find the $O(\Lambda _{\rm QCD}\left/E_{\gamma}\right.)$ contribution is also very important, but not as important as the $B$ meson decay.

\appendix

\section{\label{sec:ap1}the details of $\Phi _{\bar{q}}^{(1)}\otimes T^{(0)}$ and $\Phi _Q^{(1)}\otimes T^{(0)}$}

The calculation of $\Phi _q^{(1)}\otimes T_a^{(0)}$, $\Phi _q^{(1)}\otimes T_b^{(0)}$, $\Phi _Q^{(1)}\otimes T_a^{(0)}$ and $\Phi _Q^{(1)}\otimes T_b^{(0)}$ are similar as each other. There are collinear IR divergences in $F^{(1)\rm EM}_a$ and $F^{(1)\rm Weak}_b$ which are canceled by $\Phi _q^{(1)}\otimes T_a^{(0)}$ and $\Phi _q^{(1)}\otimes T_b^{(0)}$. The IR Div in $F^{(1)\rm EM}_a$ can be canceled to all orders, which has been proved in Ref.~\cite{Our work}. On the other hand, the IR Div in $F^{(1)\rm Weak}_b$ becomes sub-leading order when $E_{\gamma}$ is small, which is new, and can not be canceled if we use $T_a^{(0)}$ in Ref.~\cite{Our work}. So the calculation of $\Phi _q^{(1)}\otimes T_b^{(0)}$, and the cancellation of the IR Div is vital. We establish the detail of the calculation of $\Phi _q^{(1)}\otimes T_b^{(0)}$ as an example of how those corrections of wave-function are calculated. In the calculation we use Integral by part(IBP) method~\cite{IBP}.

The tree-level amplitude $T_b^{(0)}$ is
\begin{equation}
\begin{split}
&T_b^{(0)}=e_QP_L^{\mu}\frac{\slashed k_Q-\slashed p_{\gamma}+m_Q}{(k_Q-p_{\gamma})^2-m_Q^2}\slashed \varepsilon\\
\end{split}
\label{eq.a1}
\end{equation}
Then the convolution of the wave-function and the hard amplitude can be written as
\begin{equation}
\begin{split}
&\Phi_q^{(1)}\otimes T^{(0)}_{b}=ie_Qg_s^2C_F\int\frac{d^d l}{(2\pi)^d}\frac{1}{l^2}\bar{v}_{\bar{q}}\gamma ^{\rho}\frac{\slashed l + \slashed p_{\bar{q}}}{(l+p_{\bar{q}})^2-m_q^2}\int_0^1 d\alpha \left(-\left.\frac{\partial T_{b}^{(0)\mu}}{\partial k_Q^{\rho}}\right|_{k_Q=p_Q-\alpha l} \right)u_Q\\
&=-ie_Qg_s^2C_F\int\frac{d^d l}{(2\pi)^d}\frac{1}{l^2}\bar{v}_{\bar{q}}\gamma ^{\rho}\frac{\slashed l + \slashed p_{\bar{q}}}{(l+p_{\bar{q}})^2-m_q^2}P_L^{\mu}\\
&\times \int_0^1 d\alpha \left(\frac{\gamma _{\rho}}{(Q-\alpha l)^2-m_Q^2} - \frac{2 (Q-\alpha l )^{\rho}\left(\slashed Q -\alpha \slashed l  +m_Q\right)}{((Q-\alpha l)^2-m_Q^2)^2} \right)\slashed \varepsilon u_Q\\
&Q=p_Q-p_{\gamma}\\
\end{split}
\label{eq.a2}
\end{equation}
With the help of $\bar{v}_{\bar{q}}$, we find
\begin{equation}
\begin{split}
&\Phi_q^{(1)}\otimes T^{(0)}_{b}=-e_Qg_s^2C_F\bar{v}_{\bar{q}}\left(I_1+I_2+I_3\right)\slashed \varepsilon u_Q\\
\end{split}
\label{eq.a3}
\end{equation}
with
\begin{equation}
\begin{split}
&I_1=2\int\frac{d^d l}{(2\pi)^d}\frac{1}{l^2}\frac{1}{(l+p_{\bar{q}})^2-m_q^2}P_L^{\mu}\int_0^1 d\alpha \frac{\slashed p_{\bar{q}}}{(Q-\alpha l)^2-m_Q^2}\\
&I_2=-2\int\frac{d^d l}{(2\pi)^d}\frac{1}{l^2}\frac{1}{(l+p_{\bar{q}})^2-m_q^2}P_L^{\mu}\int_0^1 d\alpha \frac{(2 Q\cdot p_{\bar{q}}-\alpha 2 l\cdot p_{\bar{q}})\left(\slashed Q -\alpha \slashed l  +m_Q\right)}{((Q-\alpha l)^2-m_Q^2)^2} \\
&I_3=\int\frac{d^d l}{(2\pi)^d}\frac{1}{l^2}\frac{\gamma ^{\rho} \slashed l}{(l+p_{\bar{q}})^2-m_q^2}P_L^{\mu}
\int_0^1 d\alpha \left(\frac{\gamma _{\rho}}{(Q-\alpha l)^2-m_Q^2} - \frac{2 (Q-\alpha l )^{\rho}\left(\slashed Q -\alpha \slashed l  +m_Q\right)}{((Q-\alpha l)^2-m_Q^2)^2} \right)\\
\end{split}
\label{eq.a4}
\end{equation}
And $I_1$ can be treated as
\begin{equation}
\begin{split}
&I_1=2\int\frac{d^d l}{(2\pi)^d}\frac{1}{l^2}\int_0^1 d\alpha\frac{1}{(l+p_{\bar{q}})^2-m_q^2}P_L^{\mu} \frac{(Q^2-m_Q^2-\alpha ^2 l^2 + 2\alpha ^2l^2 - 2\alpha l\cdot Q)\slashed p_{\bar{q}}}{((Q-\alpha l)^2-m_Q^2)^2}\\
&=2\int\frac{d^d l}{(2\pi)^d}\int_0^1 d\alpha\frac{1}{l^2}\frac{1}{(l+p_{\bar{q}})^2-m_q^2}P_L^{\mu} \frac{(- 2\alpha l\cdot Q)\slashed p_{\bar{q}}}{((Q-\alpha l)^2-m_Q^2)^2}\\
&+2\int\frac{d^d l}{(2\pi)^d}\int_0^1 d\alpha \frac{2\alpha ^2}{(l+p_{\bar{q}})^2-m_q^2}P_L^{\mu}\frac{\slashed p_{\bar{q}}}{((Q-\alpha l)^2-m_Q^2)^2}\\
&+2\int\frac{d^d l}{(2\pi)^d}\frac{1}{l^2}\frac{1}{(l+p_{\bar{q}})^2-m_q^2}P_L^{\mu}\frac{\slashed p_{\bar{q}}}{((Q-\alpha l)^2-m_Q^2)^2}\\
\end{split}
\label{eq.a5}
\end{equation}
The second term in Eq.~\ref{eq.a5} can be worked out directly by integrate with $l$ first, and then integrate with $\alpha$. The result is free from IR Div, for convenience, we define
\begin{equation}
\begin{split}
&f_1=\int\frac{d^d l}{(2\pi)^d}\int_0^1 d\alpha \frac{2\alpha ^2}{(l+p_{\bar{q}})^2-m_q^2}\frac{1}{((Q-\alpha l)^2-m_Q^2)^2}\\
&=-2\frac{i}{16\pi^2}\int _0^1 dx\int _0^1 d\alpha \frac{1}{(x-1)Q^2+m_Q^2-\alpha (1-x)2Q\cdot p_{\bar{q}}-i\epsilon}\\
\end{split}
\label{eq.f1}
\end{equation}
From the integral we find $f_1$ is free from IR Div. The other terms will be calculated with $I_2$ together later. Part of $I_2$ can be worked out, and we define
\begin{equation}
\begin{split}
&f_2=2\int\frac{d^d l}{(2\pi)^d}\frac{1}{l^2}\frac{1}{(l+p_{\bar{q}})^2-m_q^2}\int_0^1 d\alpha \frac{\alpha 2 l\cdot p_{\bar{q}}\left(\slashed Q -\alpha \slashed l  +m_Q\right)}{((Q-\alpha l)^2-m_Q^2)^2} \\
&=2\int\frac{d^d l}{(2\pi)^d}\left(\frac{1}{l^2}-\frac{1}{l+2l\cdot p_{\bar{q}}}\right)\int_0^1 d\alpha \frac{\alpha \slashed Q -\alpha^2 \slashed l  +\alpha m_Q}{((Q-\alpha l)^2-m_Q^2)^2} \\
\end{split}
\label{eq.a6}
\end{equation}
When integrate with $\alpha$, the divergences in $\frac{1}{l^2}$ and $\frac{1}{l+2l\cdot p_{\bar{q}}}$ in Eq.~(\ref{eq.a6}) can cancel each other, and we find
\begin{equation}
\begin{split}
&f_2=2\frac{-i}{16\pi^2}\int_0^1 d\alpha \int_0^1 dx \frac{1}{\alpha} \left(\frac{x(1-x)\slashed Q +x m_Q}{xm_Q^2+x(x-1)Q^2-i\epsilon}\right.\\
&\left.-\frac{x(1-x)\slashed Q + \alpha x (1-x)\slashed p_{\bar{q}}+x m_Q}{\alpha x (x-1) 2Q\cdot p_{\bar{q}} + x m_Q^2 +x(x-1)Q^2-i\epsilon}\right)\\
&=-2\frac{i}{16\pi^2}\int_0^1 d\alpha \int_0^1 dx \left(\frac{((x-1)2Q\cdot p_{\bar{q}})\left((1-x)\slashed Q + m_Q\right)}{\left(m_Q^2+(x-1)Q^2-i\epsilon\right)\left(\alpha  (x-1) 2Q\cdot p_{\bar{q}} +  m_Q^2 +(x-1)Q^2-i\epsilon\right)}\right.\\
&\left.-\frac{ (1-x)\slashed p_{\bar{q}}}{\alpha  (x-1) 2Q\cdot p_{\bar{q}} +  m_Q^2 +(x-1)Q^2-i\epsilon}\right)\\
\end{split}
\label{eq.a7}
\end{equation}
From the integral we find $f_2$ is free from IR Div. The other part of $I_2$ can be written as
\begin{equation}
\begin{split}
&-2\int\frac{d^d l}{(2\pi)^d}\frac{1}{l^2}\frac{1}{(l+p_{\bar{q}})^2-m_q^2}P_L^{\mu}\int_0^1 d\alpha \frac{2 Q\cdot p_{\bar{q}}\left(\slashed Q +m_Q\right)}{((Q-\alpha l)^2-m_Q^2)^2}\\
&=-2\int\frac{d^d l}{(2\pi)^d}\frac{1}{l^2}\frac{1}{(l+p_{\bar{q}})^2-m_q^2}P_L^{\mu}\int_0^1 d\alpha \frac{(Q^2-m_Q^2)-\alpha ^2 l^2}{Q^2-m_Q^2}\frac{2 Q\cdot p_{\bar{q}}\left(\slashed Q +m_Q\right)}{((Q-\alpha l)^2-m_Q^2)^2}\\
&-2\int\frac{d^d l}{(2\pi)^d}\frac{1}{(l+p_{\bar{q}})^2-m_q^2}P_L^{\mu}\int_0^1 d\alpha \frac{\alpha ^2}{Q^2-m_Q^2}\frac{2 Q\cdot p_{\bar{q}}\left(\slashed Q +m_Q\right)}{((Q-\alpha l)^2-m_Q^2)^2}\\
&=-2P_L^{\mu}\frac{2Q\cdot p_{\bar{q}}\left(\slashed Q +m_Q\right)}{Q^2-m_Q^2}C_0+f_1 \frac{w-z}{y}P_L^{\mu}\left(\slashed Q +m_Q\right)\\
\end{split}
\label{eq.a8}
\end{equation}
where $C_0$ is the Pa-Ve function $C_0(-z,x-y+w-z,x;0,0,x)$. We will use $C$ denote $C(-z,x-y+w-z,x;0,0,x)$ in this section. We find
\begin{equation}
\begin{split}
&I_1+I_2=P_L^{\mu}f_1\left(2\slashed p_{\bar{q}}+\frac{w-y}{y}\left(\slashed Q +m_Q\right)\right)+P_L^{\mu}f_2+\left(2P_L^{\mu}\slashed p_{\bar{q}}-2P_L^{\mu}\frac{2Q\cdot p_{\bar{q}}\left(\slashed Q +m_Q\right)}{Q^2-m_Q^2}\right)C_0\\
&+2\int\frac{d^d l}{(2\pi)^d}\frac{1}{l^2}\frac{1}{(l+p_{\bar{q}})^2-m_q^2}P_L^{\mu}\int_0^1 d\alpha \frac{2Q\cdot p_{\bar{q}}\alpha \slashed l -2\alpha l\cdot Q \slashed p_{\bar{q}}}{((Q-\alpha l)^2-m_Q^2)^2}\\
\end{split}
\label{eq.a9}
\end{equation}
We can perform the Feynman parameter directly on the last term, so that
\begin{equation}
\begin{split}
&2\int\frac{d^d l}{(2\pi)^d}\frac{1}{l^2}\frac{1}{(l+p_{\bar{q}})^2-m_q^2}P_L^{\mu}\int_0^1 d\alpha \frac{2Q\cdot p_{\bar{q}}\alpha \slashed l -2\alpha l\cdot Q \slashed p_{\bar{q}}}{((Q-\alpha l)^2-m_Q^2)^2}\\
&=2\frac{1}{Q^2}\int\frac{d^d l}{(2\pi)^d}\frac{1}{l^2}\frac{1}{(l+p_{\bar{q}})^2-m_q^2}P_L^{\mu}\\
&\times \int_0^1 d\alpha \frac{Q^2(2Q\cdot p_{\bar{q}}\alpha \slashed l -2\alpha l\cdot Q \slashed p_{\bar{q}})-(Q^2-m_Q^2)(2Q\cdot p_{\bar{q}}\slashed Q- 2Q^2 \slashed p_{\bar{q}})}{((Q-\alpha l)^2-m_Q^2)^2}\\
&+2\frac{1}{Q^2}\int\frac{d^d l}{(2\pi)^d}\frac{1}{l^2}\frac{1}{(l+p_{\bar{q}})^2-m_q^2}P_L^{\mu}\int_0^1 d\alpha \frac{(Q^2-m_Q^2)(2Q\cdot p_{\bar{q}}\slashed Q- 2Q^2 \slashed p_{\bar{q}})}{((Q-\alpha l)^2-m_Q^2)^2}\\
\end{split}
\label{eq.a10}
\end{equation}
The first term is
\begin{equation}
\begin{split}
&2\frac{P_L^{\mu}(2Q\cdot p_{\bar{q}}\slashed Q- 2Q^2 \slashed p_{\bar{q}})}{Q^2}\frac{i}{16\pi^2} \int _0^1d\alpha \int _0^1 dx\int _0^{1-x}dy \frac{x^2Q^2-xq^2+xm_Q^2}{(\alpha ^2 y^2 m_q^2+\alpha x y 2Q\cdot p_{\bar{q}}+x(x-1)Q^2+xm_Q^2)^2}\\
\end{split}
\label{eq.a11}
\end{equation}
We add an $m_q^2$ term in the numerator and integrate with $\alpha$, so, it is
\begin{equation}
\begin{split}
&2\frac{P_L^{\mu}(2Q\cdot p_{\bar{q}}\slashed Q- 2Q^2 \slashed p_{\bar{q}})}{Q^2}\frac{i}{16\pi^2} \int _0^1d\alpha \int _0^1 dx\int _0^{1-x}dy \frac{ x^2Q^2-xq^2+xm_Q^2-\alpha ^2 y^2 m_q^2}{(\alpha ^2 y^2 m_q^2+\alpha x y 2Q\cdot p_{\bar{q}}+x(x-1)Q^2+xm_Q^2)^2}\\
&=2\frac{P_L^{\mu}(2Q\cdot p_{\bar{q}}\slashed Q- 2Q^2 \slashed p_{\bar{q}})}{Q^2}\frac{i}{16\pi^2} \int _0^1 dx\int _0^{1-x}dy \frac{1}{(y^2 m_q^2+ x y 2Q\cdot p_{\bar{q}}+x(x-1)Q^2+xm_Q^2)}\\
&=-2\frac{P_L^{\mu}(2Q\cdot p_{\bar{q}}\slashed Q- 2Q^2 \slashed p_{\bar{q}})}{Q^2} C_0\\
\end{split}
\label{eq.a12}
\end{equation}
And the second term is
\begin{equation}
\begin{split}
&2\frac{1}{Q^2}\int\frac{d^d l}{(2\pi)^d}\frac{1}{l^2}\frac{1}{(l+p_{\bar{q}})^2-m_q^2}P_L^{\mu}\int_0^1 d\alpha \frac{(Q^2-m_Q^2)(2Q\cdot p_{\bar{q}}\slashed Q- 2Q^2 \slashed p_{\bar{q}})}{((Q-\alpha l)^2-m_Q^2)^2}\\
&=2\frac{1}{Q^2}\int\frac{d^d l}{(2\pi)^d}\frac{1}{l^2}\frac{1}{(l+p_{\bar{q}})^2-m_q^2}P_L^{\mu}\int_0^1 d\alpha \frac{(Q^2-m_Q^2-\alpha ^2 l^2)(2Q\cdot p_{\bar{q}}\slashed Q- 2Q^2 \slashed p_{\bar{q}})}{((Q-\alpha l)^2-m_Q^2)^2}\\
&+2\frac{1}{Q^2}\int\frac{d^d l}{(2\pi)^d}\frac{1}{(l+p_{\bar{q}})^2-m_q^2}P_L^{\mu}\int_0^1 d\alpha \frac{\alpha ^2(2Q\cdot p_{\bar{q}}\slashed Q- 2Q^2 \slashed p_{\bar{q}})}{((Q-\alpha l)^2-m_Q^2)^2}\\
\end{split}
\label{eq.a13}
\end{equation}
The first term can be integrated with $\alpha$ directly, and the second term is just $f_1$, so it is
\begin{equation}
\begin{split}
&2\frac{P_L^{\mu}(2Q\cdot p_{\bar{q}}\slashed Q- 2Q^2 \slashed p_{\bar{q}})}{Q^2}C_0+f_1\frac{P_L^{\mu}(2Q\cdot p_{\bar{q}}\slashed Q- 2Q^2 \slashed p_{\bar{q}})}{Q^2}
\end{split}
\label{eq.a14}
\end{equation}
Finally, we find
\begin{equation}
\begin{split}
&I_1+I_2=P_L^{\mu}f_1\left(\left(\frac{w-z}{y}+\frac{w-z}{x-y}\right)\slashed Q + \frac{w-z}{y} m_Q\right)+P_L^{\mu}f_2\\
&+\left(2P_L^{\mu}\slashed p_{\bar{q}}-2P_L^{\mu}\frac{2Q\cdot p_{\bar{q}}\left(\slashed Q +m_Q\right)}{Q^2-m_Q^2}\right)C_0\\
\end{split}
\label{eq.a15}
\end{equation}
The $I_3$ term can be worked out together with $I_4$, which we defined as
\begin{equation}
\begin{split}
&I_4=\int\frac{d^d l}{(2\pi)^d}\frac{1}{l^2}\frac{\slashed l \gamma ^{\rho} }{(l+p_{\bar{q}})^2-m_q^2}P_L^{\mu}
\int_0^1 d\alpha \left(\frac{\gamma _{\rho}}{(Q-\alpha l)^2-m_Q^2} - \frac{2 (Q-\alpha l )^{\rho}\left(\slashed Q -\alpha \slashed l  +m_Q\right)}{((Q-\alpha l)^2-m_Q^2)^2} \right)\\
\end{split}
\label{eq.a16}
\end{equation}
the sum of $I_3$ and $I_4$ is
\begin{equation}
\begin{split}
&I_3+I_4=2\int\frac{d^d l}{(2\pi)^d}\frac{1}{l^2}\frac{1}{(l+p_{\bar{q}})^2-m_q^2}P_L^{\mu}\\
&\times \int_0^1 d\alpha \left(\frac{\slashed l}{(Q-\alpha l)^2-m_Q^2} - \frac{(2l\cdot Q - 2\alpha l^2)\left(\slashed Q -\alpha \slashed l  +m_Q\right)}{((Q-\alpha l)^2-m_Q^2)^2} \right)\\
&=-2\int\frac{d^d l}{(2\pi)^d}\frac{1}{l^2}\frac{1}{(l+p_{\bar{q}})^2-m_q^2}P_L^{\mu}
\int_0^1 d\alpha \frac{\partial }{\partial \alpha} \frac{\slashed Q - \alpha \slashed l +m_Q}{(Q-\alpha l)^2 - m_Q^2}\\
&=2\int\frac{d^d l}{(2\pi)^d}\frac{1}{l^2}\frac{1}{(l+p_{\bar{q}})^2-m_q^2}P_L^{\mu}
\left(\frac{\slashed Q +m_Q}{Q^2 - m_Q^2}-\frac{\slashed Q -\slashed l +m_Q}{(Q-l)^2 - m_Q^2}\right)\\
&=2\int\frac{d^d l}{(2\pi)^d}\frac{1}{l^2}\frac{1}{(l+p_{\bar{q}})^2-m_q^2}P_L^{\mu}
\left(-\frac{\slashed Q +m_Q}{Q^2-m_Q^2}\frac{2l\cdot Q}{(Q-l)^2-m_Q^2}+\frac{\slashed l}{(Q-l)^2 - m_Q^2}\right)\\
&+2\frac{1}{Q^2-m_Q^2}\int\frac{d^d l}{(2\pi)^d}\frac{1}{(l+p_{\bar{q}})^2-m_q^2}P_L^{\mu}\frac{\slashed Q +m_Q}{(Q-l)^2-m_Q^2}\\
\end{split}
\label{eq.a17}
\end{equation}
Using the IBP relation
\begin{equation}
\begin{split}
&0=-I_3+\int \frac{d^D l}{(2\pi)^D}\int _0^1d\alpha \frac{D}{l^2}\frac{1}{(l+p_{\bar{q}})^2-m_q^2}P_L^{\mu} \left(\frac{1}{\alpha}\frac{\slashed Q -\alpha \slashed l + m_Q}{(Q-\alpha l)^2-m_Q^2}\right)\\
&-2\int \frac{d^D l}{(2\pi)^D}\int _0^1d\alpha \frac{1}{l^2}\frac{1}{(l+p_{\bar{q}})^2-m_q^2}P_L^{\mu} \left(\frac{1}{\alpha}\frac{\slashed Q -\alpha \slashed l + m_Q}{(Q-\alpha l)^2-m_Q^2}\right)\\
&-2\int \frac{d^D l}{(2\pi)^D}\int _0^1d\alpha \frac{(\slashed l + \slashed p_{\bar{q}})\slashed l}{l^2}\frac{1}{((l+p_{\bar{q}})^2-m_q^2)^2}P_L^{\mu} \left(\frac{1}{\alpha}\frac{\slashed Q -\alpha \slashed l + m_Q}{(Q-\alpha l)^2-m_Q^2}\right)\\
\end{split}
\label{eq.a18}
\end{equation}
and
\begin{equation}
\begin{split}
&0=-I_4+\int \frac{d^D l}{(2\pi)^D}\int _0^1d\alpha \frac{D}{l^2}\frac{1}{(l+p_{\bar{q}})^2-m_q^2}P_L^{\mu} \left(\frac{1}{\alpha}\frac{\slashed Q -\alpha \slashed l + m_Q}{(Q-\alpha l)^2-m_Q^2}\right)\\
&-2\int \frac{d^D l}{(2\pi)^D}\int _0^1d\alpha \frac{1}{l^2}\frac{1}{(l+p_{\bar{q}})^2-m_q^2}P_L^{\mu} \left(\frac{1}{\alpha}\frac{\slashed Q -\alpha \slashed l + m_Q}{(Q-\alpha l)^2-m_Q^2}\right)\\
&-2\int \frac{d^D l}{(2\pi)^D}\int _0^1d\alpha \frac{\slashed l(\slashed l + \slashed p_{\bar{q}})}{l^2}\frac{1}{((l+p_{\bar{q}})^2-m_q^2)^2}P_L^{\mu} \left(\frac{1}{\alpha}\frac{\slashed Q -\alpha \slashed l + m_Q}{(Q-\alpha l)^2-m_Q^2}\right)\\
\end{split}
\label{eq.a19}
\end{equation}
we find
\begin{equation}
\begin{split}
&I_3-I_4=2\int\frac{d^d l}{(2\pi)^d} \int _0^1 d\alpha  \frac{\slashed l \slashed p_{\bar{q}}-\slashed l \slashed p_{\bar{q}}}{l^2}\frac{1}{((l+p_{\bar{q}})^2-m_q^2)^2}P_L^{\mu} \left(\frac{1}{\alpha}\frac{\slashed Q -\alpha \slashed l + m_Q}{(Q-\alpha l)^2-m_Q^2}\right)\\
\end{split}
\label{eq.a20}
\end{equation}
With the help of $\bar{v}_{\bar{q}}$, it is
\begin{equation}
\begin{split}
&I_3-I_4=2\int\frac{d^d l}{(2\pi)^d} \int _0^1 d\alpha  \frac{2l\cdot p_{\bar{q}}}{l^2}\frac{1}{((l+p_{\bar{q}})^2-m_q^2)^2}P_L^{\mu} \left(\frac{1}{\alpha}\frac{\slashed Q -\alpha \slashed l + m_Q}{(Q-\alpha l)^2-m_Q^2}\right)\\
\end{split}
\label{eq.a21}
\end{equation}
The $\alpha$ integral can be worked out by the substitution $l\to \alpha l$
\begin{equation}
\begin{split}
&I_3-I_4=2\int\frac{d^d (\alpha l)}{(2\pi)^d} \int _0^1 d\alpha  \frac{2(\alpha l)\cdot p_{\bar{q}}}{(\alpha l)^2}\frac{1}{((\alpha l)^2+\alpha 2(\alpha l)\cdot p_{\bar{q}})^2}P_L^{\mu} \frac{\slashed Q -\alpha \slashed l + m_Q}{(Q-\alpha l)^2-m_Q^2}\\
&=2\int\frac{d^d l}{(2\pi)^d} \int _0^1 d\alpha  \frac{2l\cdot p_{\bar{q}}}{l^2}\frac{1}{(l^2+\alpha 2l\cdot p_{\bar{q}})^2}P_L^{\mu} \frac{\slashed Q - \slashed l + m_Q}{(Q-l)^2-m_Q^2}\\
\end{split}
\label{eq.a22}
\end{equation}
And now we can integrate with $\alpha$, and the result is
\begin{equation}
\begin{split}
&I_3-I_4=2\int\frac{d^d l}{(2\pi)^d} \frac{2l\cdot p_{\bar{q}}}{l^4}\frac{1}{l^2+2l\cdot p_{\bar{q}}}P_L^{\mu} \frac{\slashed Q+ m_Q}{(Q-l)^2-m_Q^2}
 +2\int\frac{d^d l}{(2\pi)^d} \frac{1}{l^4}P_L^{\mu} \frac{- \slashed l }{(Q-l)^2-m_Q^2}
 +2\slashed C\\
\end{split}
\label{eq.a23}
\end{equation}
where $C^{\mu}$ is the Pa-Ve function. And then we find
\begin{equation}
\begin{split}
&2\int\frac{d^d l}{(2\pi)^d} \frac{2l\cdot p_{\bar{q}}}{l^4}\frac{1}{l^2+2l\cdot p_{\bar{q}}}P_L^{\mu} \frac{\slashed Q + m_Q}{(Q-l)^2-m_Q^2}\\
&=2\int\frac{d^d l}{(2\pi)^d} \frac{2l\cdot p_{\bar{q}}}{l^4}\frac{1}{l^2+2l\cdot p_{\bar{q}}}P_L^{\mu} \frac{\slashed Q + m_Q}{Q^2-m_Q^2}\\
&+2\frac{1}{Q^2-m_Q^2}\int\frac{d^d l}{(2\pi)^d} \frac{2l\cdot p_{\bar{q}}}{l^4}\frac{1}{l^2+2l\cdot p_{\bar{q}}}P_L^{\mu} \frac{2l\cdot Q(\slashed Q + m_Q)}{(Q-l)^2-m_Q^2}\\
&-2\frac{1}{Q^2-m_Q^2}\int\frac{d^d l}{(2\pi)^d} \frac{2l\cdot p_{\bar{q}}}{l^2}\frac{1}{l^2+2l\cdot p_{\bar{q}}}P_L^{\mu} \frac{\slashed Q + m_Q}{(Q-l)^2-m_Q^2}\\
\end{split}
\label{eq.a24}
\end{equation}
The first term will vanish. Then we can calculate $I_3$ by using $2I_3=(I_3+I_4)+(I_3-I_4)$, so
\begin{equation}
\begin{split}
&I_3=\int\frac{d^d l}{(2\pi)^d}\frac{1}{l^2}\frac{1}{(l+p_{\bar{q}})^2-m_q^2}P_L^{\mu}
\left(-2\frac{\slashed Q +m_Q}{Q^2-m_Q^2}\frac{2l\cdot Q}{(Q-l)^2-m_Q^2}+2\frac{\slashed l}{(Q-l)^2 - m_Q^2}\right)\\
&+2\frac{1}{Q^2-m_Q^2}\int\frac{d^d l}{(2\pi)^d}\frac{1}{(l+p_{\bar{q}})^2-m_q^2}P_L^{\mu}\frac{\slashed Q +m_Q}{(Q-l)^2-m_Q^2}\\
&+\frac{1}{Q^2-m_Q^2}\int\frac{d^d l}{(2\pi)^d} \frac{1}{l^4}P_L^{\mu} \frac{2l\cdot Q(\slashed Q + m_Q)}{(Q-l)^2-m_Q^2}\\
&-\frac{1}{Q^2-m_Q^2}\int\frac{d^d l}{(2\pi)^d} \frac{1}{l^2}P_L^{\mu} \frac{\slashed Q + m_Q}{(Q-l)^2-m_Q^2}
 +\int\frac{d^d l}{(2\pi)^d} \frac{1}{l^4}P_L^{\mu} \frac{- \slashed l }{(Q-l)^2-m_Q^2}\\
\end{split}
\label{eq.a25}
\end{equation}

Using the results
\begin{equation}
\begin{split}
&\frac{1}{Q^2-m_Q^2}\int\frac{d^d l}{(2\pi)^d} \frac{1}{l^4}P_L^{\mu} \frac{2l\cdot Q(\slashed Q + m_Q)}{(Q-l)^2-m_Q^2}
=-2\frac{1}{y}\frac{i}{16\pi^2}\left(1-\frac{x}{x-y}\log \frac{x}{y}\right)P_L^{\mu}\left(\slashed Q + m_Q\right)\\
&\int\frac{d^d l}{(2\pi)^d} \frac{1}{l^4}P_L^{\mu} \frac{- \slashed l}{(Q-l)^2-m_Q^2}
=-\frac{i}{16\pi^2}\frac{(x-y)-x\log \frac{x}{y}}{(x-y)^2}P_L^{\mu} \slashed Q\\
\end{split}
\label{eq.a26}
\end{equation}
we define
\begin{equation}
\begin{split}
&f_3=-\frac{i}{16\pi^2}\left(\left(\frac{2}{y}-\frac{2x}{y(x-y)}\log \frac{x}{y}\right) (\slashed Q+m_Q)+\left(\frac{1}{x-y}-\frac{x}{(x-y)^2}\log \frac{x}{y}\right)\slashed Q\right)\\
\end{split}
\label{eq.a27}
\end{equation}
So that, combine all those terms together we have
\begin{equation}
\begin{split}
&I_1+I_2+I_3
=P_L^{\mu}f_1\left(\left(\frac{w-z}{y}+\frac{w-z}{x-y}\right)\slashed Q + \frac{w-z}{y} m_Q\right)+P_L^{\mu}f_2+P_L^{\mu}f_3\\
&-\frac{2P_L^{\mu}(\slashed Q+m_Q)}{y}B_0(x-y+w-z,0,x)+\frac{P_L^{\mu}(\slashed Q+m_Q)}{y}B_0(x-y,0,x)\\
&+\left(2P_L^{\mu}\slashed p_{\bar{q}}-2P_L^{\mu}\frac{2Q\cdot p_{\bar{q}}\left(\slashed Q +m_Q\right)}{Q^2-m_Q^2}\right)C_0
+2P_L^{\mu}\left(-\frac{\slashed Q +m_Q}{Q^2-m_Q^2}2Q\cdot C+\slashed C\right)\\
\end{split}
\label{eq.a28}
\end{equation}

Now we are going to show the IR divergence cancellation. $F_b^{(1)\rm Weak}$ can be written as
\begin{equation}
\begin{split}
&F_{b}^{(1)\rm Weak}=ie_QC_fg_s^2\int \frac{d^dl}{(2\pi)^d}\bar{v}_{\bar{q}}\frac{1}{l^2}\gamma _{\rho}\frac{\slashed l+\slashed p_{\bar{q}}}{(l+p_{\bar{q}})^2-m_q^2}P_L^{\mu}\frac{\slashed Q-\slashed l+m_Q}{(Q-l)^2-m_Q^2}\gamma ^{\rho}\frac{\slashed Q+m_Q}{Q^2-m_Q^2}\slashed \varepsilon u_Q\\
&=ie_QC_fg_s^2\bar{v}(F_1+F_2+F_3+F_4)\slashed \varepsilon u_Q\\
\end{split}
\label{eq.a29}
\end{equation}
where
\begin{equation}
\begin{split}
&F_1=\frac{1}{l^2}\gamma _{\rho}\frac{\slashed l}{(l+p_{\bar{q}})^2-m_q^2}P_L^{\mu}\frac{-\slashed l}{(Q-l)^2-m_Q^2}\gamma ^{\rho}\frac{\slashed Q+m_Q}{Q^2-m_Q^2}\\
&F_2=\frac{1}{l^2}\gamma _{\rho}\frac{\slashed p_{\bar{q}}}{(l+p_{\bar{q}})^2-m_q^2}P_L^{\mu}\frac{\slashed Q+m_Q}{(Q-l)^2-m_Q^2}\gamma ^{\rho}\frac{\slashed Q+m_Q}{Q^2-m_Q^2}\\
&F_3=\frac{1}{l^2}\gamma _{\rho}\frac{\slashed l}{(l+p_{\bar{q}})^2-m_q^2}P_L^{\mu}\frac{\slashed Q+m_Q}{(Q-l)^2-m_Q^2}\gamma ^{\rho}\frac{\slashed Q+m_Q}{Q^2-m_Q^2}\\
&F_4=\frac{1}{l^2}\gamma _{\rho}\frac{\slashed p_{\bar{q}}}{(l+p_{\bar{q}})^2-m_q^2}P_L^{\mu}\frac{-\slashed l}{(Q-l)^2-m_Q^2}\gamma ^{\rho}\frac{\slashed Q+m_Q}{Q^2-m_Q^2}\\
\end{split}
\label{eq.a30}
\end{equation}
with the help of $\bar{v}_{\bar{q}}$, they are
\begin{equation}
\begin{split}
&F_1=(D-2)\left(-B_0(x-y+w-z,0,x)\right)P_L^{\mu}\frac{\slashed Q+m_Q}{Q^2-m_Q^2}\\
&+2(D-2)C_{00}P_L^{\mu}\frac{\slashed Q+m_Q}{Q^2-m_Q^2}-4(p_{\bar{q}}^{\mu}C_{12}-Q^{\mu}C_{22})(1+\gamma _5)\frac{\slashed Q m_Q+Q^2}{Q^2-m_Q^2}\\
&F_2=2P_L^{\mu}\left(-\slashed p_{\bar{q}}+\frac{2Q\cdot p_{\bar{q}}(\slashed Q + m_Q)}{Q^2-m_Q^2}\right)C_0\\
&F_3=\left(-2Q^2P_L^{\mu}\frac{\slashed Q+m_Q}{Q^2-m_Q^2}+2P_L^{\mu}\slashed Q+2\slashed Q\gamma^{\mu}(1+\gamma _5)\right)C_2+2P_L^{\mu} \left(-\slashed p_{\bar{q}}+\frac{2Q\cdot p_{\bar{q}}(\slashed Q + m_Q)}{Q^2-m_Q^2}\right)C_1\\
&F_4=\left(2P_L^{\mu}2Q\cdot p_{\bar{q}}\frac{\slashed Q +m_Q}{Q^2-m_Q^2}-4p_{\bar{q}}^{\mu}(1+\gamma _5)\frac{Q^2+\slashed Q m_Q}{Q^2-m_Q^2}\right)C_2\\
\end{split}
\label{eq.a31}
\end{equation}
Finally, we can obtain
\begin{equation}
\begin{split}
&\Phi ^{(0)}\otimes T_b^{(1)\;{\rm Weak}}=ie_Qg_s^2C_F\bar{v}_{\bar{q}}\\
&\times \left(\left(\frac{y-w+z}{w-z}\left(B_0(x-y,0,x)-B_0(x-y+w-z,0,x)\right)+2\right)P_L^{\mu}\frac{\slashed Q+m_Q}{Q^2-m_Q^2}\right.\\
&\left.+2\slashed Q\gamma^{\mu}(1+\gamma _5)C_2-4p_{\bar{q}}^{\mu}(1+\gamma _5)\frac{Q^2+\slashed Q m_Q}{Q^2-m_Q^2}C_2-4(p_{\bar{q}}^{\mu}C_{12}-Q^{\mu}C_{22})(1+\gamma _5)\frac{\slashed Q m_Q+Q^2}{Q^2-m_Q^2}\right.\\
&\left.+P_L^{\mu}f_1\left(\left(\frac{w-z}{y}+\frac{w-z}{x-y}\right)\slashed Q + \frac{w-z}{y} m_Q\right)+P_L^{\mu}f_2+P_L^{\mu}f_3\right)\slashed \varepsilon u_Q\\
\end{split}
\label{eq.a32}
\end{equation}
We know that, there are IR divergences in both, $C_0$ and $C_1$. However, they are canceled and will not show up in $T_b^{(1){\rm Weak}}$, and the cancellation can be shown without using any order expansion over the heavy quark mass. Therefor, at 1-loop order, the cancellation of IR Div can be established to all orders over the heavy quark mass expansion.

\section{\label{sec:ap2}the detail of $\Phi _{\rm Wfc}^{(1)} \otimes T^{(0)}$}
The correction of wave-function corresponding to $F_a^{(1)\rm Wfc}$ and $F_b^{(1)\rm Wfc}$ is $\Phi _{\rm Wfc}^{(1)} \otimes T_a^{(0)}$ and $\Phi _{\rm Wfc}^{(1)} \otimes T_b^{(0)}$. The calculation of these two corrections are similar to each other.

The $\Phi _{\rm Wfc}^{(1)} \otimes T_a^{(0)}$ is
\begin{equation}
\begin{split}
&\Phi^{(1)}_{\rm Wfc}\otimes T_a^{(0)}=(D-2)ig_s^2C_F\int\frac{d^D l}{(2\pi)^D}\frac{1}{l^2}\bar{v}_{\bar{q}}\int _0^1d\alpha\int _0^1d\beta \frac{\slashed q - (\alpha -\beta)\slashed l}{(q-(\alpha -\beta)l)^4} u_Q\\
\end{split}
\label{b1}
\end{equation}
At first we adjust the region of $\alpha$ and $\beta$, and we find
\begin{equation}
\begin{split}
&\int\frac{d^D l}{(2\pi)^D}\int _0^1d\alpha \int _0^1 d\beta \frac{1}{l^2} \frac{\slashed q-(\alpha-\beta)\slashed l}{(q-(\alpha -\beta)l)^4}\\
&=\int\frac{d^D l}{(2\pi)^D}\int _0^1d\beta \int _{-\beta}^{1-\beta} d\alpha \frac{1}{l^2} \frac{\slashed q-\alpha \slashed l}{(q-\alpha l)^4}\\
&=\int\frac{d^D l}{(2\pi)^D}\left(\int _{-1}^0d\alpha \int _{-\alpha}^1d\beta +\int _0^1 d\alpha \int _0^{1-\alpha}d\beta \right)\frac{1}{l^2} \frac{\slashed q-\alpha \slashed l}{(q-\alpha l)^4}\\
&=\int\frac{d^D l}{(2\pi)^D}\left(\int _{-1}^0d\alpha (1+\alpha) +\int _0^1 d\alpha (1-\alpha) \right)\frac{1}{l^2} \frac{\slashed q-\alpha \slashed l}{(q-\alpha l)^4}\\
&=\int\frac{d^D l}{(2\pi)^D}\int _{-1}^1d\alpha \frac{1}{l^2} \frac{\slashed q-\alpha \slashed l}{(q-\alpha l)^4}\\
&+\int\frac{d^D l}{(2\pi)^D}\left(\int _{-1}^0d\alpha \alpha -\int _0^1 d\alpha \alpha \right)\frac{1}{l^2} \frac{\slashed q-\alpha \slashed l}{(q-\alpha l)^4}\\
\end{split}
\label{b2}
\end{equation}

We calculate the first term at first, we shift the momentum $l\to l-\frac{q}{\alpha}$, and we find
\begin{equation}
\begin{split}
\int\frac{d^D l}{(2\pi)^D}\frac{1}{(q-\alpha l)^4} \frac{\slashed l}{l^2}=-\int\frac{d^D l}{(2\pi)^D}\frac{1}{\alpha} \frac{1}{(q+\alpha l)^2} \frac{\slashed l}{l^4}
\end{split}
\label{b3}
\end{equation}
Using IBP relation
\begin{equation}
\begin{split}
&0=(D-3)\int\frac{d^D l}{(2\pi)^D}\frac{1}{\alpha} \frac{1}{(q+\alpha l)^2} \frac{\slashed l}{l^4}-\int\frac{d^D l}{(2\pi)^D}\frac{(2\alpha l^2 + 2l\cdot q)}{(q+\alpha l)^4}\frac{\slashed l}{l^4}\\
\end{split}
\label{b4}
\end{equation}
we find
\begin{equation}
\begin{split}
&\int _{-1}^1 d\alpha\int \frac{d^D l}{(2\pi)^D}\frac{1}{\alpha} \frac{1}{(q+\alpha l)^2} \frac{\slashed l}{l^4}=\int \frac{d^D l}{(2\pi)^D} \left(\frac{1}{(q-l)^2}\frac{\slashed l}{l^4}-\frac{1}{(q+l)^2}\frac{\slashed l}{l^4}\right)=2\frac{\slashed q}{q^2}\frac{i}{16\pi^2}
\end{split}
\label{b5}
\end{equation}

Then we calculate the second term. We can obtain
\begin{equation}
\begin{split}
&\int \frac{d^D l}{(2\pi)^D}\left(\int _{-1}^0d\alpha \alpha -\int _0^1 d\alpha \alpha\right)\frac{\slashed q -\alpha \slashed l}{(q-\alpha l)^4}\frac{1}{l^2}\\
&=2\int _0^1dX\int \frac{d^D l}{(2\pi)^D}\left(\int _{-1}^0d\alpha \alpha -\int _0^1 d\alpha \alpha \right)\frac{X(1-X)\slashed q}{\alpha ^4(l^2-\frac{x(1-x)(-q^2)}{\alpha^2})^3}\\
&=-\int _0^1dX\int \frac{d^D l}{(2\pi)^D}\frac{\slashed q}{q^2}\frac{1}{(l^2-X(1-X)(-q^2))^2}=-\frac{\slashed q}{q^2}\frac{i}{16\pi^2}\left(N_{\rm UV}-\log \frac{z}{\mu^2}+2\right)\\
\end{split}
\label{b6}
\end{equation}
Finally, we find
\begin{equation}
\begin{split}
&\Phi^{(1)}_{\rm Wfc}\otimes T_a^{(0)}=-2ig_s^2C_F\bar{v}_{\bar{q}}\slashed \varepsilon \frac{\slashed q}{q^2}P_L^{\mu}u_Q\frac{i}{16\pi^2}\left(N_{\rm UV}-\log \frac{z}{\mu^2}+3\right)\\
\end{split}
\label{b7}
\end{equation}

Next we calculate $\Phi _{\rm wfc}^{(1)} \otimes T_b^{(0)}$. The correction can be written as
\begin{equation}
\begin{split}
&\Phi^{(1)\rm Wfc}\otimes T^{(0)}_{b}=-i2g_s^2e_QC_f\int\frac{d^D l}{(2\pi)^D}\frac{1}{l^2}\bar{v}_{\bar{q}}P_L^{\mu}\int _0^1d\alpha\int _0^1d\beta \left(\frac{2m_Q^2(\slashed Q +(\alpha -\beta)\slashed l+m_Q)}{((Q+(\alpha -\beta)l)^2-m_Q^2)^3}\right.\\
&\left.-(\frac{D-2}{2})\frac{(\slashed Q +(\alpha -\beta)\slashed l)}{((Q+(\alpha -\beta)l)^2-m_Q^2)^2}-(\frac{D-4}{2})\frac{m_Q}{((Q+(\alpha -\beta)l)^2-m_Q^2)^2}\right)\slashed \varepsilon u_Q\\
&Q=p_Q-p_{\gamma}\\
\end{split}
\label{b8}
\end{equation}
At first, we adjust the region of $\alpha$ and $\beta$, first, we make the substitution $\alpha \to \alpha -\beta$, then we change the order of $\alpha$ and $\beta$, so
\begin{equation}
\begin{split}
&I=\int _0^1d\alpha\int _0^1d\beta \left(\frac{2m_Q^2(\slashed Q +(\alpha -\beta)\slashed l+m_Q)}{((Q+(\alpha -\beta)l)^2-m_Q^2)^3}\right.\\
&\left.-(\frac{D-2}{2})\frac{(\slashed Q +(\alpha -\beta)\slashed l)}{((q+\alpha )l)^2-m_Q^2)^2}-(\frac{D-4}{2})\frac{m_Q}{((Q+\alpha l)^2-m_Q^2)^2}\right)\frac{1}{l^2}\\
&=\int _{-\beta}^{1-\beta}d\alpha\int _0^1d\beta \left(\frac{2m_Q^2(\slashed Q +\alpha \slashed l+m_Q)}{((Q+\alpha l)^2-m_Q^2)^3}\right.\\
&\left.-(\frac{D-2}{2})\frac{(\slashed Q +\alpha \slashed l)}{((Q+\alpha l)^2-m_Q^2)^2}-(\frac{D-4}{2})\frac{m_Q}{((Q+\alpha l)^2-m_Q^2)^2}\right)\frac{1}{l^2}\\
&=\left(\int _{-1}^0d\alpha \int _{-\alpha}^1 d\beta+\int _0^1 d\alpha \int _0^{1-\alpha}d\beta \right)\left(\frac{2m_Q^2(\slashed Q +\alpha \slashed l+m_Q)}{((Q+\alpha l)^2-m_Q^2)^3}\right.\\
&\left.-(\frac{D-2}{2})\frac{(\slashed Q +\alpha \slashed l)}{((Q+\alpha l)^2-m_Q^2)^2}-(\frac{D-4}{2})\frac{m_Q}{((Q+\alpha l)^2-m_Q^2)^2}\right)\frac{1}{l^2}\\
\end{split}
\label{b9}
\end{equation}
After integrating over the parameter $\beta$, we can obtain
\begin{equation}
\begin{split}
&I=I_1+I_2\\
&I_1=\int _{-1}^1d\alpha\left(\frac{2m_Q^2(\slashed Q +\alpha\slashed l+m_Q)}{((Q+\alpha l)^2-m_Q^2)^3}\right.\\
&\left.-(\frac{D-2}{2})\frac{(\slashed Q +\alpha\slashed l)}{((Q+\alpha l)^2-m_Q^2)^2}-(\frac{D-4}{2})\frac{m_Q}{((Q+\alpha l)^2-m_Q^2)^2}\right)\frac{1}{l^2}\\
&I_2=\left(\int _{-1}^0d\alpha \alpha -\int _0^1 d\alpha \alpha\right)\left(\frac{2m_Q^2(\slashed Q +\alpha \slashed l+m_Q)}{((Q+\alpha l)^2-m_Q^2)^3}\right.\\
&\left.-(\frac{D-2}{2})\frac{(\slashed Q +\alpha \slashed l)}{((Q+\alpha l)^2-m_Q^2)^2}-(\frac{D-4}{2})\frac{m_Q}{((Q+\alpha l)^2-m_Q^2)^2}\right)\frac{1}{l^2}\\
\end{split}
\label{b10}
\end{equation}

Define
\begin{equation}
\begin{split}
&i_1=\int _{-1}^1d\alpha \int \frac{d^D l}{(2\pi)^D}\frac{\slashed Q}{((Q+\alpha l)^2-m_Q^2)^2}\frac{1}{l^2}\\
&=-\int _{-1}^1d\alpha \int \frac{d^D l}{(2\pi)^D}\frac{\slashed Q}{Q^2-m_Q^2}\frac{\alpha ^2 l^2 -(Q^2-m_Q^2)}{((Q+\alpha l)^2-m_Q^2)^2}\frac{1}{l^2}\\
&+\int _{-1}^1d\alpha \int \frac{d^D l}{(2\pi)^D}\frac{\slashed Q}{Q^2-m_Q^2}\frac{\alpha ^2 l^2}{((Q+\alpha l)^2-m_Q^2)^2}\frac{1}{l^2}\\
\end{split}
\label{b11}
\end{equation}
the integrating over $\alpha$ in the first term can be integrated directly, and using IBP relation
\begin{equation}
\begin{split}
&0=\int \frac{d^D l}{(2\pi)^D}\left(Q\cdot \frac{\partial }{\partial l}\frac{\alpha ^2 \slashed l}{((Q+\alpha l)^2-m_Q^2)^2}\right)\\
\end{split}
\label{b12}
\end{equation}
we can obtain
\begin{equation}
\begin{split}
&\int \frac{d^D l}{(2\pi)^D}\frac{\alpha ^2 \slashed Q}{((Q+\alpha l)^2-m_Q^2)^2}\\
&=2\int \frac{d^D l}{(2\pi)^D}\frac{(2\alpha ^2 l\cdot Q+2\alpha (Q^2-m_Q^2))\alpha ^2\slashed l}{((Q+\alpha l)^2-m_Q^2)^3}+2\int \frac{d^D l}{(2\pi)^D}\frac{(2\alpha m_Q^2 \slashed l)\alpha ^2}{((Q+\alpha l)^2-m_Q^2)^3}\\
\end{split}
\label{b13}
\end{equation}
The integrating over $\alpha$ in the first term in Eq.~\ref{b13} can be integrated directly, and the second term can be written as
\begin{equation}
\begin{split}
&\int \frac{d^D l}{(2\pi)^D}\frac{(2\alpha m_Q^2 \slashed l)\alpha ^2}{((Q+\alpha l)^2-m_Q^2)^3}=-\int \frac{d^D l}{(2\pi)^D}\frac{2 m_Q^2 \alpha^2\slashed Q}{((Q+\alpha l)^2-m_Q^2)^3}\\
\end{split}
\label{b14}
\end{equation}
Then using IBP relation
\begin{equation}
\begin{split}
&0=D\int \frac{d^D l}{(2\pi)^D}\frac{\alpha^2\slashed Q}{((Q+\alpha l)^2-m_Q^2)^3}-3\int \frac{d^D l}{(2\pi)^D}\frac{\left(2\alpha ^2 l^2+2\alpha l\cdot Q\right)\alpha^2\slashed Q}{((Q+\alpha l)^2-m_Q^2)^4}\\
\end{split}
\label{b15}
\end{equation}
we obtain
\begin{equation}
\begin{split}
&\int \frac{d^D l}{(2\pi)^D}\frac{\alpha^2\slashed Q}{((Q+\alpha l)^2-m_Q^2)^3}=\frac{3}{D-3}\int \frac{d^D l}{(2\pi)^D}\frac{\left(\alpha ^2 l^2-Q^2+m_Q^2\right)\alpha^2\slashed Q}{((Q+\alpha l)^2-m_Q^2)^4}\\
\end{split}
\label{b16}
\end{equation}
For simplicity, we define the integration of Eq.~\ref{b16} over $\alpha$, i.e.
\begin{equation}
\begin{split}
&i_2=\int _{-1}^1d\alpha \int \frac{d^D l}{(2\pi)^D}\frac{\alpha^2\slashed Q}{((Q+\alpha l)^2-m_Q^2)^3}=\frac{i}{16\pi^2}\frac{\slashed Q}{m_Q^2}\\
\end{split}
\label{b17}
\end{equation}
Then we can get the result of $i_1$
\begin{equation}
\begin{split}
&i_1=\int \frac{d^D l}{(2\pi)^D} \frac{1}{Q^2-m_Q^2}\left(\frac{\slashed Q}{(Q+l)^2-m_Q^2}\frac{1}{l^2}+\frac{\slashed Q}{(Q-l)^2-m_Q^2}\frac{1}{l^2}\right.\\
&\left.+\frac{\slashed l}{((Q+l)^2-m_Q^2)^2}-\frac{\slashed l}{((Q-l)^2-m_Q^2)^2}\right)-4\frac{1}{Q^2-m_Q^2}m_Q^2i_2\\
&=\frac{i}{16\pi^2}\frac{\slashed Q}{Q^2-m_Q^2}\left(-2\frac{y}{x-y}\log\frac{x}{y}\right)=2\frac{i}{16\pi^2}\frac{\slashed Q}{x-y}\log\frac{x}{y}\\
\end{split}
\label{b18}
\end{equation}
and use IBP relation we find
\begin{equation}
\begin{split}
&0=(D-1)\int \frac{d^D l}{(2\pi)^D}\frac{\alpha \slashed l}{((Q+\alpha l)^2-m_Q^2)^2}\frac{1}{l^2}-2\int \frac{d^D l}{(2\pi)^D}\frac{(2\alpha ^2 l^2 +2\alpha l\cdot Q)\alpha \slashed l}{((Q+\alpha l)^2-m_Q^2)^3}\frac{1}{l^2}\\
\end{split}
\label{b19}
\end{equation}
we define it as $i_3$, and we find
\begin{equation}
\begin{split}
&i_3=\int \frac{d^D l}{(2\pi)^D}\frac{\alpha \slashed l}{((Q+\alpha l)^2-m_Q^2)^2}\frac{1}{l^2}=\int \frac{d^D l}{(2\pi)^D}\frac{1}{D-3}\frac{(2 \alpha ^3 l^2 - 2\alpha  (Q^2-m_Q^2))\slashed l}{((Q+\alpha l)^2-m_Q^2)^3}\frac{1}{l^2}\\
\end{split}
\label{b20}
\end{equation}
the $\alpha$ can be integrated out directly, so we find
\begin{equation}
\begin{split}
&i_3=-2\frac{i}{16\pi^2}\frac{\slashed Q}{x-y}\left(1+\log\frac{x}{y}-\frac{x}{x-y}\log\frac{x}{y}\right)\\
\end{split}
\label{b21}
\end{equation}
And we find that $\lim _{\substack {m_Q\to 0}}\left(i_2+i_3\right)$ can reproduce the result of integral in the light quark case as expected.

And then is
\begin{equation}
\begin{split}
&\int _{-1}^1 d\alpha\int \frac{d^D l}{(2\pi)^D}2m_Q^2\frac{(\slashed Q +\alpha \slashed l+m_Q)}{((Q+\alpha l)^2-m_Q^2)^3}\frac{1}{l^2}
\end{split}
\label{b22}
\end{equation}
we find
\begin{equation}
\begin{split}
&2m_Q^2\frac{(\slashed Q+m_Q)}{((Q+\alpha l)^2-m_Q^2)^3}\frac{1}{l^2}=2m_Q^2\left(1+\frac{\slashed Q m_Q}{Q^2}\right)\frac{\slashed Q}{((Q+\alpha l)^2-m_Q^2)^3}\frac{1}{l^2}\\
\end{split}
\label{b23}
\end{equation}
so we define
\begin{equation}
\begin{split}
&i_4=\int _{-1}^1 d\alpha\int \frac{d^D l}{(2\pi)^D}\frac{\slashed Q}{((Q+\alpha l)^2-m_Q^2)^3}\frac{1}{l^2}\\
&=\frac{1}{Q^2-m_Q^2}\int _{-1}^1 d\alpha\int \frac{d^D l}{(2\pi)^D}\left(\frac{\slashed Q}{((Q+\alpha l)^2-m_Q^2)^2}\frac{1}{l^2}\right.\\
&\left.+\frac{\alpha ^2 l^2 \slashed Q}{((Q+\alpha l)^2-m_Q^2)^3}\frac{1}{l^2}-\frac{(2\alpha ^2 l^2 +2\alpha l\cdot Q )\slashed Q}{((Q+\alpha l)^2-m_Q^2)^3}\frac{1}{l^2}\right)\\
\end{split}
\label{b24}
\end{equation}
the first term is just $i_1$, the second term is just $i_2$, and the third term can be transformed to $i_1$ use IBP relation
\begin{equation}
\begin{split}
&0=(D-2)\int \frac{d^D l}{(2\pi)^D}\frac{\slashed Q}{((Q+\alpha l)^2-m_Q^2)^2}\frac{1}{l^2}-2\int \frac{d^D l}{(2\pi)^D}\frac{(2\alpha l\cdot Q +2\alpha ^2 l^2)\slashed Q}{((Q+\alpha l)^2-m_Q^2)^3}\frac{1}{l^2}\\
\end{split}
\label{b25}
\end{equation}
so we find
\begin{equation}
\begin{split}
&i_4=-\frac{1}{y}\left((\frac{D-2}{2}+1)i_1+i_2\right)\\
\end{split}
\label{b26}
\end{equation}
And then we define
\begin{equation}
\begin{split}
&i_5=\int _{-1}^1 d\alpha\int \frac{d^D l}{(2\pi)^D}\frac{\alpha \slashed l}{((Q+\alpha l)^2-m_Q^2)^3}\frac{1}{l^2}
\end{split}
\label{b27}
\end{equation}
The calculation of $i_4$ is simpler then $i_2$, we have
\begin{equation}
\begin{split}
&i_5=\frac{1}{Q^2-m_Q^2}\int _{-1}^1 d\alpha\int \frac{d^D l}{(2\pi)^D}\left(\frac{\alpha \slashed l}{((Q+\alpha l)^2-m_Q^2)^2}\frac{1}{l^2}-\frac{1}{2}\frac{(2\alpha l^2 + 2l\cdot Q)\alpha \slashed l}{((Q+\alpha l)^2-m_Q^2)^3}\frac{1}{l^2}\right.\\
&\left.-\frac{1}{2}\frac{2\alpha l\cdot Q \alpha \slashed l}{((Q+\alpha l)^2-m_Q^2)^3}\frac{1}{l^2}\right)\\
\end{split}
\label{b28}
\end{equation}
the first term is $i_3$, and also use IBP relation, we find the second term can be transformed to $i_3$, as
\begin{equation}
\begin{split}
&0=(D-1)\int \frac{d^D l}{(2\pi)^D}\frac{\alpha \slashed l}{((Q+\alpha l)^2-m_Q^2)^2}\frac{1}{l^2}-2\int \frac{d^D l}{(2\pi)^D}\frac{(2\alpha l\cdot Q +2\alpha ^2 l^2)\alpha \slashed l}{((Q+\alpha l)^2-m_Q^2)^3}\frac{1}{l^2}\\
\end{split}
\label{b29}
\end{equation}
there is no UV Div in $i_3$, so use $D=4$ we find
\begin{equation}
\begin{split}
&i_5=\frac{1}{Q^2-m_Q^2}\left(\frac{i_3}{4}+\int _{-1}^1 d\alpha\int \frac{d^D l}{(2\pi)^D}\left(-\frac{1}{2}\frac{2\alpha l\cdot Q \alpha \slashed l}{((Q+\alpha l)^2-m_Q^2)^3}\frac{1}{l^2}\right)\right)\\
\end{split}
\label{b30}
\end{equation}
use IBP relation
\begin{equation}
\begin{split}
&0=D\int \frac{d^D l}{(2\pi)^D}\frac{2\alpha^2 l\cdot Q \slashed l}{((Q+\alpha l)^2-m_Q^2)^3}\frac{1}{l^2}-3\int \frac{d^D l}{(2\pi)^D}\frac{(2\alpha ^2 l^2 +2\alpha l\cdot Q)2\alpha^2 l\cdot Q \slashed l}{((Q+\alpha l)^2-m_Q^2)^4}\frac{1}{l^2}\\
&=(D-3)\int \frac{d^D l}{(2\pi)^D}\frac{2\alpha^2 l\cdot Q \slashed l}{((Q+\alpha l)^2-m_Q^2)^3}\frac{1}{l^2}-3\int \frac{d^D l}{(2\pi)^D}\frac{(\alpha ^2 l^2 -Q^2+m_Q^2)2\alpha^2 l\cdot Q \slashed l}{((Q+\alpha l)^2-m_Q^2)^4}\frac{1}{l^2}\\
\end{split}
\label{b31}
\end{equation}
the $\alpha$ can be integrated now, and finally, we find
\begin{equation}
\begin{split}
&i_5=-\frac{i}{16\pi^2}\frac{1}{(x-y)}\frac{i}{16\pi^2}\left(\frac{1}{x}-\frac{\log \frac{x}{y}}{ (x-y)}\right)\slashed Q\\
\end{split}
\label{b32}
\end{equation}
and the last term is
\begin{equation}
\begin{split}
&\int _{-1}^1d\alpha \int \frac{d^D l}{(2\pi)^D}\frac{m_Q}{((Q+\alpha l)^2-m_Q^2)^2}\frac{1}{l^2}=\frac{\slashed Q m_Q}{Q^2}i_1\\
\end{split}
\label{b33}
\end{equation}
And now we can write $I_1$ as
\begin{equation}
\begin{split}
&I_1=\frac{i}{16\pi^2}\left(-\frac{\slashed Q+m_Q}{y}\left(\frac{8x}{x-y}\log\frac{x}{y}+2\right)\right)\\
\end{split}
\label{b34}
\end{equation}

And then is the integral with $I_2$, after Feynman parameter, and neglect the terms with even times of $\alpha$ because they will vanish, and then we find
\begin{equation}
\begin{split}
&I_2=\int _0^1 dX \left(\int _{-1}^0d\alpha \alpha -\int _0^1 d\alpha \alpha \right)\int \frac{d^Dl}{(2\pi)^D}\left(\frac{\Gamma (4)X^2}{\Gamma (3)}\frac{2m_Q^2((1-X)\slashed Q +m_Q)}{\alpha ^6(l^2-\frac{\Delta}{\alpha^2})^4}\right.\\
&\left.-(\frac{D-2}{2})\frac{\Gamma (3)X}{\Gamma (2)}\frac{((1-X)\slashed Q)}{\alpha ^4(l^2-\frac{\Delta}{\alpha^2})^3}-(\frac{D-4}{2})\frac{\Gamma (3)X}{\Gamma (2)}\frac{m_Q}{\alpha ^4(l^2-\frac{\Delta}{\alpha^2})^3}\right)\\
&\Delta = X^2 Q^2-XQ^2+Xm_Q^2\\
\end{split}
\label{b35}
\end{equation}
Now we can integrate $\alpha$ directly, so
\begin{equation}
\begin{split}
&I_2=-\int _0^1 dX \int \frac{d^Dl}{(2\pi)^D} \left(\frac{\Gamma (4)X^2}{\Gamma (3)}\frac{l^2-3\Delta}{6 \Delta^2}\frac{2m_Q^2((1-X)\slashed Q +m_Q)}{(l^2-\Delta)^3}\right.\\
&\left.-(\frac{D-2}{2})\frac{\Gamma (3)X}{\Gamma (2)}\frac{1}{-2\Delta}\frac{((1-X)\slashed Q)}{(l^2-\Delta)^2}-(\frac{D-4}{2})\frac{\Gamma (3)X}{\Gamma (2)}\frac{1}{-2\Delta}\frac{m_Q}{(l^2-\Delta)^2}\right)\\
&=-\frac{i}{16\pi^2}\left(\frac{\slashed Q + m_Q}{y}\left(N_{\rm UV}-\log\frac{y}{\mu^2}+\frac{x}{x-y}\log\frac{x}{y}\right)+\frac{\slashed Q}{(x-y)^2} \left(y-x+x \log\frac{x}{y}\right)\right)\\
\end{split}
\label{b36}
\end{equation}
Finally, we find
\begin{equation}
\begin{split}
&\Phi^{(1)\rm Wfc}\otimes T^{(0)}_{b}=-i2g_s^2C_F\frac{i}{16\pi^2}\bar{v}_{\bar{q}}P_L^{\mu}\left(\frac{m_Q}{y}\left(\frac{x}{x-y}\log\frac{x}{y}+\frac{y}{x-y}-\frac{x^2}{(x-y)^2}\log\frac{x}{y}\right)\right.\\
&\left.+\frac{\slashed Q+m_Q}{y}\left(-N_{\rm UV}+\log\frac{y}{\mu^2}-\frac{8x}{x-y}\log\frac{x}{y}+\frac{y}{x-y}-\frac{x^2 \log \frac{x}{y}}{(x-y)^2}-2\right)\right)\slashed \varepsilon u_Q\\
\end{split}
\label{b37}
\end{equation}
Use $m_Q\to 0$, we can see it will reproduce the light quark result.

%%%%%%%%%%%%%%%%%%%%%%%%%%%%%%%%%%%%%%%%%%%%%%%%%%%%%%%%%%%%%%%%%%%%%%%%%
% ACKNOWLEDGMENTS
%%%%%%%%%%%%%%%%%%%%%%%%%%%%%%%%%%%%%%%%%%%%%%%%%%%%%%%%%%%%%%%%%%%%%%%%%

\textbf{Acknowledgements}: This work is supported in part by the
National Natural Science Foundation of China under contracts Nos.
11375088, 10975077, 10735080, 11125525.

\end{document}